# Dynamic Analysis of Centralized Energy Storage Systems — A Comparison between Grid-following and Grid-forming Controls

Qiang Fu [a], Siqi Bu [a], Yang Wang [b], and Mingyu Yan [c]

[a] Department of Electrical and Electronic Engineering, The Hong Kong Polytechnic University, Hong Kong, China.

[b] College of Electrical Engineering, Sichuan University, Chengdu, China.

[c] School of Electrical and Electronics Engineering, Huazhong University of Science and Technology, Wuhan, China.

*Abstract*—**This study investigates the small-signal stability of centralized energy storage systems (CESSs) using grid-following (GFL) and grid-forming (GFM) controls, particularly focusing on bidirectional power flow and multiple energy storage systems (ESSs). To address the issue of complex dynamics in CESSs when comprehensive GFL and GFM control loops are considered, high-order dynamics are simplified using the virtual damping method by focusing on the dominant oscillation mode. Damping analysis verifies that CESSs using a single-type control (either GFL or GFM) have dynamic superimposition characteristics. Specifically, as ESS number increases, the damping of GFM-CESSs improves but that of GFL-CESSs decreases. The damping sensitivity shows that the damping of GFM-CESSs is more sensitive to bidirectional power flow and all control loops, whereas that of GFL-CESSs is more sensitive to *d*-axis control loop. Consequently, GFM-CESSs are preferred for large-scale integration but are limited in scenarios with significant power reversal. If GFL and GFM controls are hybridized in CESSs, the ratio of GFM-CESSs should be constrained to avoid instability from modal resonance between GFL-CESSs and GFM-CESSs. This highlights that implementing GFM-CESSs necessitates considering scenario limitations rather than pursuing maximal integration under hybrid integration conditions. The conclusions are validated through modal analysis and time-domain simulations.**

## I. Introduction

Energy storage systems (ESSs) have gained much attention due to their functional capabilities, such as power balance [1], power quality improvement [2], and oscillation mitigation [3]. With the growing penetration of renewable energy, the trend toward the application of centralized energy storage systems (CESSs) has become mainstream, as they provide better controllability and larger rated power. According to the US DOE Global Energy Storage Database, the total rated power of electrochemical batteries and chemical storage systems exceeded 20 GW in 2023 [4], demonstrating their crucial role in future power systems.

Currently, there are two major control strategies for ESSs: grid-following (GFL) control and grid-forming (GFM) control. GFL control-based ESSs (GFL-ESSs) rely on the phase-locked loop (PLL) to rapidly detect the phase information of the point of common coupling (PCC) of the connected power system. However, this approach often results in stability issues, such as subsynchronous resonance and synchronous instability. It was demonstrated in [5] that the dynamics of PLL negatively affect the ESS dynamics under weak grid connections. To address the issue that PLL may not accurately detect voltage and phase information under weak grid connections, a novel single-phase PLL technique was proposed in [6] for application under weak grid conditions. Other control parameters of GFL-ESSs may also cause instabilities. The instabilities have been investigated in [7]-[8]. If multiple GFL-ESSs are considered, the study in [9] indicated that interactions among ESSs may lead to oscillations even when each ESS operates stably. Notably, the above instability risks of GFL-ESSs are similar to those of renewable energy while operating in a power-output state. Both are caused by grid-side converters utilizing GFL control.

To completely mitigate PLL-induced instability risks, GFM control was proposed. GFM control-based ESSs (GFM-ESSs) resemble synchronous generators and do not rely on PLL [10]. Numerous studies highlight the advantages of GFM control over traditional GFL control. References [11]-[12] demonstrated that GFM control provides frequency and voltage supports, thereby improving system dynamics, making it suitable for islanded and weak power systems. In [13], a STATCOM with GFM control was verified to provide a much wider stability margin than with GFL control. In [14], a generalized impedance model was established, demonstrating that GFM control can significantly enhance external stability compared to GFL control. However, GFM control can also pose instabilities. In [15], the impact of the damping coefficient in GFM control was analyzed. The study concludes that inappropriate parameter tuning can cause oscillations. In [16]-[17], the instability risks of GFM control under strong grid connections are identified, with findings showing that active and reactive power become coupled during oscillations. Moreover, instabilities can arise from control interactions. For example, interactions between GFM controls were investigated in [18], revealing that instability occurs when the open-loop oscillation modes of two GFM controls are close. Similarly, strong interactions between GFM control and other

Qiang Fu and Siqi Bu are with the Department of Electrical and Electronic Engineering of The Hong Kong Polytechnic University. Yang Wang is with the College of Electrical Engineering of Sichuan University. Mingyu Yan is with the School of Electrical and Electronics Engineering of Huazhong University of Science and Technology.

This work is supported by the National Natural Science Foundation of China (Grant No. 52407129), and the Supported by Sichuan Science and Technology Program (Grant No. 2025YFHZ0234). We also acknowledge the support from Hong Kong Scholars Program.

Corresponding author: Siqi Bu, Email: siqi.bu@polyu.edu.hk.

devices can worsen stability [19].

It has been well known that GFL-ESSs and GFM-ESSs have different dynamics because their grid-side control structures and parameters are different. However, three critical research gaps remain for CESSs:

1) CESSs are usually simplified to a single ESS for stability analysis, such as [13]. This simplification ignores the dynamic interactions among individual ESSs and may lead to errors in stability analysis results when the scale of CESSs is large. For example, in [21], the dynamic interactions among multiple GFL-ESSs were considered, and their impact on stability was analyzed, indicating that the integration of large-scale ESSs can induce instability even when the control parameters of ESSs are well designed. Moreover, if both GFL and GFM controls are implemented in CESSs, the dynamic interactions within CESSs should be considered rather than simplified by a single ESS.

2) The bidirectional power flow characteristic of a single ESS has been widely studied. However, for CESSs including multi-ESSs, the CESS dynamics may change significantly and pose potential instabilities as the power flow direction reverses. For example, it has been concluded that the number of ESSs should be limited when ESSs operate as constant power loads. This indicates that excessive ESSs can worsen system dynamics and potentially induce instability. To address the negative resistance caused by constant power loads, a damping control strategy for ESSs was proposed in [20], which enhances positive resistance and improves system dynamics. Furthermore, if GFL and GFM controls are hybridized in CESSs, the CESS dynamics have not been investigated under bidirectional power flow.

3) The complete GFM control structure involves multiple control loops, including power synchronous control (PSC), AC voltage control (AVC), and inner current control (ICC) loops. However, most studies have focused only on a subset of these loops, which is not entirely accurate. For example, in [22], the dynamics of the virtual synchronous generator (VSG) control were analyzed, where the highest order of the GFM control model was reduced to a second-order system. When control loop coupling of GFM control is considered, the AVC loop is additionally incorporated, as shown in [23]. Furthermore, an interaction model integrating both PSC and AVC loops was established in [24], identifying critical parameters for dynamic interactions between GFM-ESSs and wind power plants.

To address the above gaps, this study compares the different impacts of bidirectional power flow and scale of CESSs on the oscillation damping, where both GFL and GFM controls are involved. The key contributions are summarized as follows:

1) CESSs with single-type control (GFL-CESSs or GFM-CESSs) show dynamic superimposition characteristic as ESS number increases. The scale of GFL-CESSs should be limited to maintain a positive damping for dominant oscillation mode, whereas the damping improves as the scale of GFM-CESSs increases. However, for CESSs with hybrid control of GFL and GFM, the ratio of GFM-CESSs needs to be limited to avoid the modal resonance between GFL-CESSs and GFM-CESSs within the CESSs.

2) It is demonstrated that bidirectional power flow affects the dynamics of GFM-CESSs and GFL-CESSs differently. The oscillation damping exhibits higher sensitivity to power flow direction when GFM-CESSs are connected, but significantly lower sensitivity when GFL-CESSs are connected. Therefore, GFM-CESSs are less suitable for applications with significant power reversal.

3) With the PSC, AVC, and ICC loops of the GFM control structure considered, the subsynchronous dynamics of GFM-CESSs are identified to be sensitive to PSC parameters and both *d*- and *q*-axis control parameters. Conversely, in GFL-CESSs, subsynchronous dynamics are primarily influenced by *d*-axis control parameters and show negligible dependence on *q*-axis control parameters. Therefore, improving oscillation damping requires different parameter adjustment priorities for GFL-CESSs and GFM-CESSs.

The remainder of this paper is organized as follows. Section II develops models of GFL-CESSs, GFM-CESSs, and hybrid CESSs. Section III investigates the dynamic superimposition of CESSs using single-type control (GFL or GFM control) under bidirectional power flow, and Section IV identifies destabilizing risks in hybrid control-based CESSs (GFL and GFM controls). Section V analyzes the potential errors caused by the involved simplifications. Section VI validates the conclusions through modal analysis, participation factor analysis, and time-domain simulations. Section VII summarizes the key findings and implications for future research.

## II. Dynamic Models of CESSs with GFL and GFM Controls

### A. Configuration of CESSs

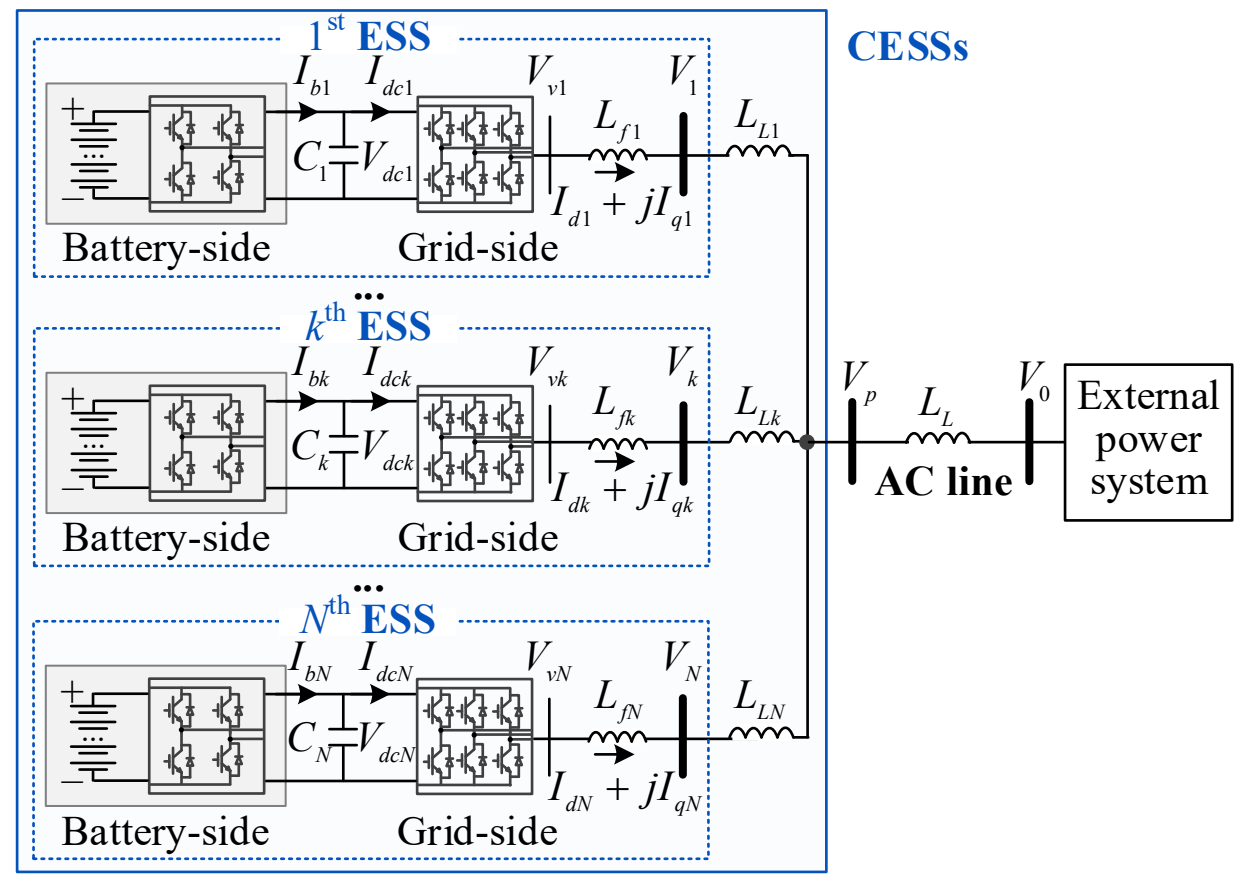

Fig. 1. Configuration of CESSs.

In Fig. 1, $N$ ESSs are centralized in parallel and linked to the external power system through an AC line. Considering that the rated power capacity of the external power system is larger than that of CESSs, the external power system is represented as an infinite bus (i.e., $\Delta V_0 = 0$) [25]. The active power exchange between the CESSs and external power system is bidirectional, depending on CESS charging/discharging states. Different from distributed ESSs, CESSs are integrated within a specific region and connected to the external power system at a single node. The ESSs inside are almost the same, so that control instructions can be sent to them consistently. For an ESS, there are two alternative controls, specifically GFL and GFM controls.

For a GFL-ESS, two mainstream control modes can be used for the grid-side converter [26]-[27]. The first is DC voltage control. In this case, active power control is adopted in the battery-side converter, and the battery-side converter can be considered a constant power source/load governed by control references [28]. The second is active power control. In this case, DC voltage control is adopted in the battery-side converter, and the battery-side converter can be simplified as a constant DC voltage source. The study in [21] demonstrates that DC voltage control of the grid-side converter exhibits more complex dynamics and is therefore the focus of this paper. The analysis process can be applied to active power control by simplifying the control structure (e.g., reducing the outer loop dynamics from second order to first order [21]). However, for a GFM-ESS, the grid-side converter needs to provide grid-forming functionalities (e.g., voltage/frequency regulation) to the AC power system. Therefore, DC voltage control needs to be adopted for the battery-side converter to ensure power balance is maintained across the DC system.

*B. Transfer Function of GFL-CESSs*

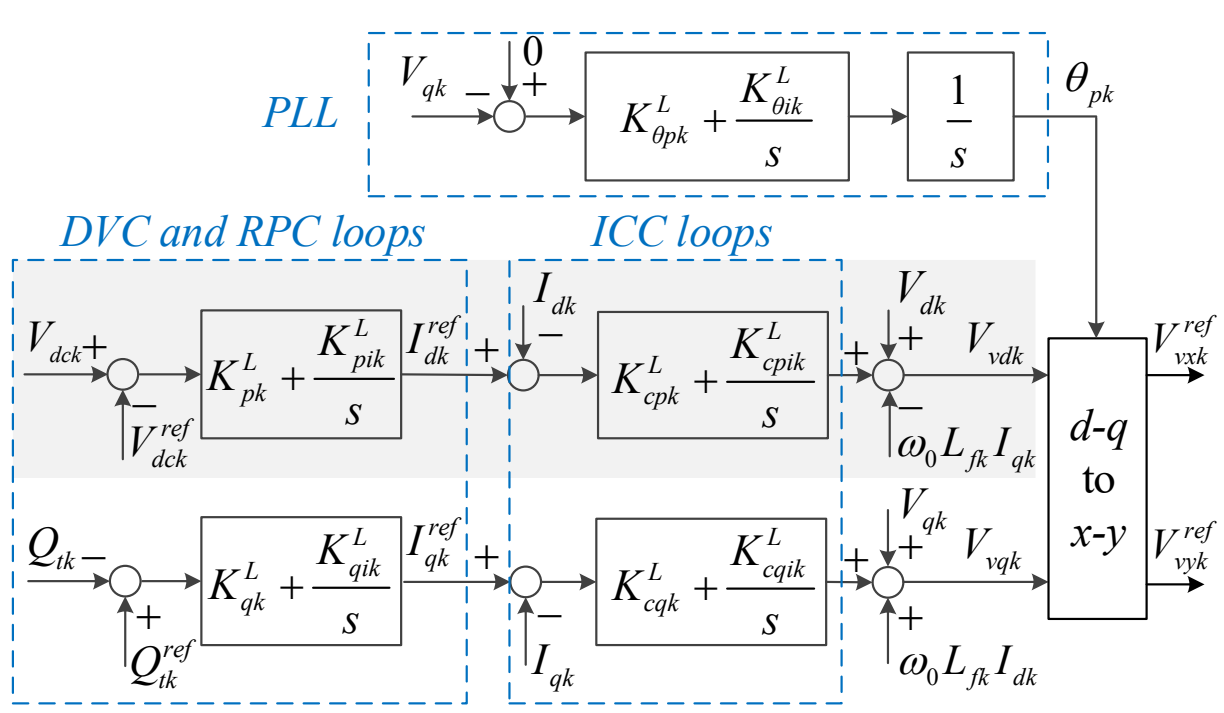

Fig. 2. GFL control of the grid-side converter in the $k^{th}$ ESS.

The configuration of grid-side control loops of the $k^{th}$ ESS is illustrated in Fig. 2, including DC voltage control (DVC) loop, reactive power control (RPC) loop, inner current control (ICC) loops, and PLL. The dynamic model of these loops has been well-documented in [29]-[30] and thus will not be detailed in this paper. The governing equations are:

The dynamic equations of the DVC loop:

$$C_k \frac{dV_{dck}}{dt} = \frac{P_{bk} - V_{dk} I_{dk}}{V_{dck}}, \quad I_{dk}^{ref} = \left( K_{pk}^L + \frac{K_{pik}^L}{s} \right) \left( V_{dck} - V_{dck}^{ref} \right) \tag{1}$$

where $C_k$ is the capacitance at the DC side of the $k^{th}$ grid-side converter, $P_{bk}$ is the active power output from the battery-side converter, $V_{dck}$ is the DC voltage of $C_k$, $I_{dk}$ is the AC current in $d$-axis, and $V_{dk}$ is the $d$-axis component of the AC voltage at the terminal of the $k^{th}$ ESS. $K_{pk}^L + K_{pik}^L s^{-1}$ is the transfer function of DVC loop of the $k^{th}$ grid-side converter.

The dynamic equation of the RPC loop:

$$I_{qk}^{ref} = \left( K_{qk}^L + \frac{K_{qik}^L}{s} \right) \left( Q_{tk}^{ref} - Q_{tk} \right), \tag{2}$$

where $Q_{tk}$ is the reactive power output from the $k^{th}$ ESS, and $I_{qk}$ is the AC current in $q$-axis. $K_{qk}^L + K_{qik}^L s^{-1}$ is the transfer function of RPC loop of the $k^{th}$ grid-side converter.

The dynamic equations of the ICC loops:

$$I_{dk} = G_{idk}^L(s) I_{dk}^{ref}, \quad I_{qk} = G_{iqk}^L(s) I_{qk}^{ref}, \tag{3}$$

where,

$$G_{idk}^L(s) = \frac{K_{cpk}^L s + K_{cpik}^L}{L_{fk} s^2 + K_{cpk}^L s + K_{cpik}^L}, \quad G_{iqk}^L(s) = \frac{K_{cqk}^L s + K_{cqik}^L}{L_{fk} s^2 + K_{cqk}^L s + K_{cqik}^L}$$

$L_{fk}$ is the inductance of the filter on the AC side of the $k^{th}$ ESS, and $\omega_0$ is the synchronous angular frequency with $\omega_0 = 1$ p.u., $K_{cpk}^L + K_{cpik}^L s^{-1}$ and $K_{cqk}^L + K_{cqik}^L s^{-1}$ are transfer functions of $d$- and $q$-axis ICC loop of the $k^{th}$ grid-side converter, respectively. Detailed derivations of (3) can be found in Appendix A.

The dynamic equations of the PLL:

$$\theta_{pk} = \left( K_{\theta pk}^L + \frac{K_{\theta ik}^L}{s} \right) \left( 0 - V_{qk} \right), \quad V_{qk} = V_k \sin\left( \theta_k - \theta_{pk} \right) \approx V_k \left( \theta_k - \theta_{pk} \right) \tag{4}$$

where $\theta_k$ and $\theta_{pk}$ are angles of $V_k$ and PLL, respectively, and $\theta_k = \theta_{pk}$ at steady states. $V_{qk}$ is the $q$-axis component of the AC voltage at the terminal of the $k^{th}$ ESS. $K_{\theta pk}^L + K_{\theta ik}^L s^{-1}$ is the transfer function of PLL of the $k^{th}$ grid-side converter.

By linearizing (1) to (4) at steady state, the transfer function of the $k^{th}$ ESS in $d$-$q$ coordinates can be obtained as

$$\begin{bmatrix} \Delta I_{dk} \\ \Delta I_{qk} \end{bmatrix} = \begin{bmatrix} H_{ddk}(s) & H_{dqk}(s) \\ H_{qdk}(s) & H_{qqk}(s) \end{bmatrix} \begin{bmatrix} \Delta V_{dk} \\ \Delta V_{qk} \end{bmatrix} = \mathbf{Y}_{dqk}^L(s) \begin{bmatrix} \Delta V_{dk} \\ \Delta V_{qk} \end{bmatrix}, \tag{5}$$

where the $d$-$q$ coordinates are rotating coordinates established on the converters, subscript 0 denotes the steady-state values of variables, and

$$\begin{aligned} H_{ddk}(s) &= -\frac{I_{dk,0} G_{idk}^L(s)(K_{pk}^L s + K_{pik}^L)}{C_k V_{dck,0} s^2 + V_{k,0} G_{idk}^L(s)(K_{pk}^L s + K_{pik}^L)} \\ H_{dqk}(s) &= -\frac{I_{qk,0} G_{idk}^L(s)(K_{pk}^L s + K_{pik}^L)}{C_k V_{dck,0} s^2 + V_{k,0} G_{idk}^L(s)(K_{pk}^L s + K_{pik}^L)} \\ H_{qdk}(s) &= -\frac{I_{qk,0} G_{iqk}^L(s)(K_{qk}^L s + K_{qik}^L)}{s + V_{k,0} G_{iqk}^L(s)(K_{qk}^L s + K_{qik}^L)} \\ H_{qqk}(s) &= \frac{I_{dk,0} G_{iqk}^L(s)(K_{qk}^L s + K_{qik}^L)}{s + V_{k,0} G_{iqk}^L(s)(K_{qk}^L s + K_{qik}^L)} \end{aligned}.$$

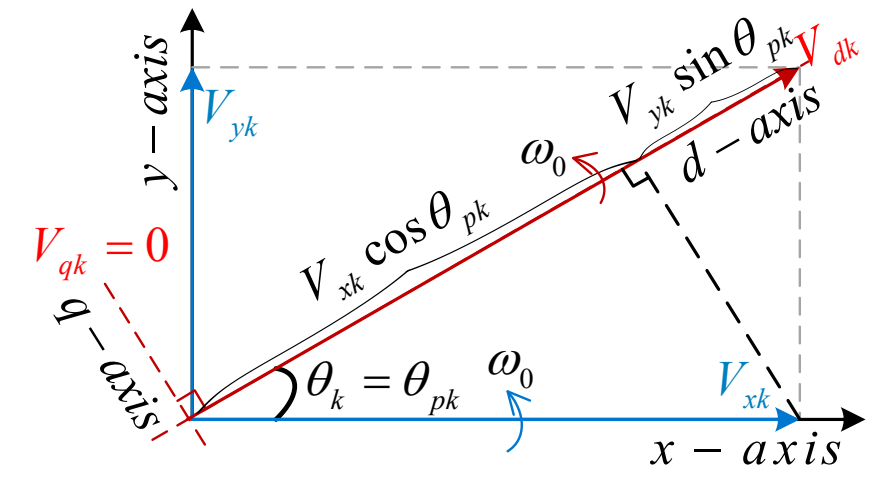

Fig. 3. Relationship between the *d-q* and *x-y* coordinates.

The transfer function of the $k^{th}$ ESS in *x-y* coordinates can be obtained as (details are in Appendix B)

$$\begin{bmatrix}\Delta I_{xk}\\ \Delta I_{yk}\end{bmatrix} = \mathbf{Y}^{L}_{xyk}(s)\begin{bmatrix}\Delta V_{xk}\\ \Delta V_{yk}\end{bmatrix}, \quad (6)$$

where the *x-y* coordinates are rotating coordinates established on the AC power system, and the relationship between the *d-q* and *x-y* coordinates is illustrated in Fig. 3 and given by (B1).

### C. Transfer Function of GFM-CESSs

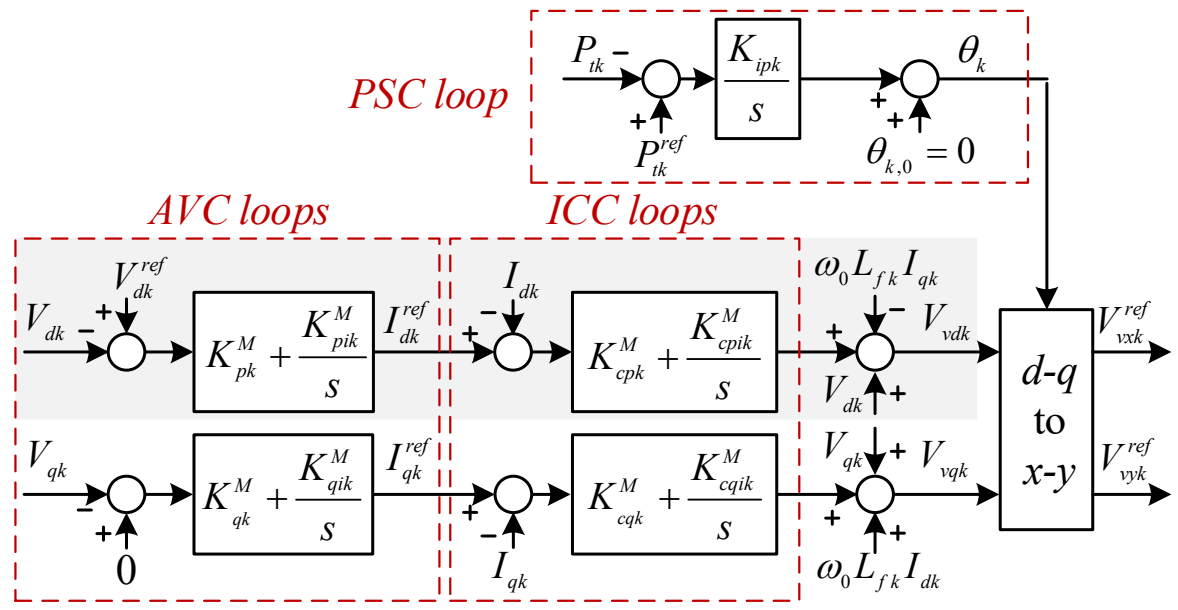

Fig. 4. GFM control of the grid-side converter in the $k^{th}$ ESS.

In Fig. 4, the AVC, ICC and PSC loops of GFM control are indicated. The relation between the *d-q* and *x-y* coordinates can be described in mathematical by (B1). The AVC loop is to control the amplitude of AC voltage at the PCC by controlling $V_{dk}$ to be the reference value, and $V_{qk}$ to be zero. The output values are as the reference values of ICC loops, the dynamic equations are obtained as:

$$\begin{aligned} I^{ref}_{dk} &= \left(K^{M}_{pk} + \frac{K^{M}_{pik}}{s}\right)\left(V^{ref}_{dk} - V_{dk}\right) \\ I^{ref}_{qk} &= \left(K^{M}_{qk} + \frac{K^{M}_{qik}}{s}\right)\left(0 - V_{qk}\right) \end{aligned}, \quad (7)$$

where $K^{M}_{pk} + K^{M}_{pik}s^{-1}$ and $K^{M}_{qk} + K^{M}_{qik}s^{-1}$ are transfer functions of AVC loops in the *d*- and *q*- axis, respectively.

The ICC loops are used to quickly lock the currents to the reference values, which is the same as that in GFL control, the dynamic equations are obtained referring to (3) as

$$\begin{bmatrix} I_{dk}\\ I_{qk}\end{bmatrix} = \begin{bmatrix} G^{M}_{idk}(s) & \\ & G^{M}_{iqk}(s)\end{bmatrix}\begin{bmatrix} I^{ref}_{dk}\\ I^{ref}_{qk}\end{bmatrix}. \quad (8)$$

The PSC loop is used to control the output power of the ESS. It's dynamics can be described by

$$\theta_k = \frac{K_{ipk}}{s}\left(P^{ref}_{tk} - P_{tk}\right) + \theta_{k,0}, \quad (9)$$

where,

$$P_{tk} = \begin{bmatrix} V_{xk} & V_{yk}\end{bmatrix}\begin{bmatrix} I_{xk}\\ I_{yk}\end{bmatrix} = \begin{bmatrix} V_{dk} & V_{qk}\end{bmatrix}\begin{bmatrix} I_{dk}\\ I_{qk}\end{bmatrix},$$

$P_{tk}$ is the active power output from the $k^{th}$ ESS. $K^{M}_{ipk}s^{-1}$ is the transfer function of PSC loop of the $k^{th}$ grid-side converter.

By linearizing (7) to (9), the transfer function of the GFM controlled converter in *x-y* coordinates can be obtained as (details are in Appendix C)

$$\begin{bmatrix}\Delta I_{xk}\\ \Delta I_{yk}\end{bmatrix} = \mathbf{Y}^{M}_{xyk}(s)\begin{bmatrix}\Delta V_{xk}\\ \Delta V_{yk}\end{bmatrix}. \quad (10)$$

### D. Transfer Function of CESSs with GFL and GFM Controls

The impedance matrix of the AC network can be written in *x-y* coordinates as

$$\mathbf{V}_{xy} = \mathbf{Z}_{net}\mathbf{I}_{xy}, \quad (11)$$

where $\mathbf{I}_{xy} = [I_{x1}, I_{y1}, \dots, I_{xN}, I_{yN}]^T$, $\mathbf{V}_{xy} = [V_{x1}, V_{y1}, \dots, V_{xN}, V_{yN}]^T$.

From (5), (10) and (11), the unified transfer function of the CESSs with different controls can be obtained to be

$$\mathbf{Z}_{net}Diag(\mathbf{Y}_{xyk}(s)) = \mathbf{E}, \quad (12)$$

where $\mathbf{Y}_{xyk}(s)$ can be $\mathbf{Y}^{L}_{xyk}(s)$, $\mathbf{Y}^{M}_{xyk}(s)$ or a hybrid combination of them. $\mathbf{E}$ is an identity matrix.

The dynamic characteristics of the CESSs can be analyzed by the eigenvalues, i.e., the solutions of (12), and the stability can be determined by the damping of eigenvalues.

## III. Damping Analysis of Dominant Oscillation Mode in CESSs with Single-Type Control

### A. Brief Introduction of Virtual Damping Analysis Method

The virtual damping analysis method [32] is effective for assessing how the second-order oscillation loop related to the dominant oscillation mode is affected by the dynamics of the remaining power system. It can quantitatively evaluate the damping contributed by the remaining dynamic loops of the power system to that second-order oscillation loop, thereby explaining why the dominant oscillation mode moves to the right half of the complex plane and causes instability from a physical damping perspective. Besides, it can reduce a high-order equation to an equivalent second-order equation while preserving accuracy around the critical stability boundary. This provides faster stability evaluation across operating conditions and avoids the time-consuming eigenvalue analysis based on the full-order state-space model.

For generality, the second-order oscillation loop associated with the dominant oscillation mode is defined as

$$T_O(s) = s^2 + bs + c, \quad (13)$$

where detailed expressions of the coefficients can be found in $F_L(s)$ of GFL control in (19), and $F_M(s)$ of GFM control in (22).

The transfer function of the remaining power system is denoted as $T_R(s)$. Accordingly, the closed-loop characteristic equation of the power system can be written as

$$T_O(s) + T_R(s) = 0. \quad (14)$$

The solutions of (14) include the dominant oscillation mode, defined as $\lambda_s = -d_s + j\omega_s$.

Considering $s = \lambda_s$, (14) equals to

$$T_O(s) + \mathrm{Re}\left[T_R(\lambda_s)\right] + \frac{d_s}{\omega_s}\mathrm{Im}\left[T_R(\lambda_s)\right] + \frac{\mathrm{Im}\left[T_R(\lambda_s)\right]}{\omega_s}s = 0. \quad (15)$$

Therefore, from (15), a higher-order characteristic equation is reduced to a second-order equation around the dominant oscillation mode. However, $\lambda_s$ is unknown, it is common to adopt the approximation $s = \lambda_s \approx j\omega_s$, because the dominant oscillation mode that may induce instability typically has very small damping. The result can accurately indicate whether the system is stable or not at $\omega_s$. If the stability over a frequency range is required to be determined, each frequency point in the range should be evaluated individually.

From (15), the damping of the dominant oscillation mode can be expressed as

$$d_s = \frac{1}{2}\left(b + \frac{\mathrm{Im}\left[T_R(\lambda_s)\right]}{\omega_s}\right). \tag{16}$$

It should be noted that the approximation $s \approx j\omega_s$ does not lead to misjudgment of stability because at the critical stability boundary, $s = j\omega_s$ holds exactly. Moreover, the variation trend of the damping obtained from (16) matches that of the actual damping.

*B. Dynamic Superimposition of CESSs with Single-type Control*

Considering that CESSs need to maintain consistency, they are usually from the same manufacturer and the parameters of each ESS are the same. Furthermore, internal lines of CESSs are significantly shorter than external lines, leading to:

$$L_L \approx L_L + L_{Li} \approx L_L + L_{Lj},\ i,j \in [1,N],\ i \neq j. \tag{17}$$

Thus, the transfer function of the power system connected with GFL-CESSs can be obtained from (12) to be

$$\mathbf{Y}^L_{xyk}(s) - \left(N\begin{bmatrix} 0 & -X_L \\ X_L & 0 \end{bmatrix}\right)^{-1} = \begin{bmatrix} 0 & 0 \\ 0 & 0 \end{bmatrix}, \tag{18}$$

where $X_L = \omega_0 L_L$, the transfer function of $\mathbf{Y}^L_{xyk}(s)$ can be simplified with $\theta_k \approx 0$, yields,

$$\mathbf{Y}^L_{xyk}(s) = \begin{bmatrix} H_{ddk}(s) & 0 \\ 0 & p_{\theta k}(s) I_{k,0} / V_{k,0} \end{bmatrix}.$$

The characteristic equation of GFL-CESSs can be obtained by calculating determinant of (18), i.e., |Eq (18)|, as

$$F_L(s) + F_{LO}(s) = 0, \tag{19}$$

where,

$$\begin{aligned} F_L(s) &= V_{k,0} C_k V_{dck,0} s^2 + V_{k,0}{}^2 G^L_{idk}(s)\left(K^L_{pk} s + K^L_{pik}\right) \\ F_{LO}(s) &= -N^2 X_L{}^2 I_{k,0}{}^2 p_{\theta k}(s) G^L_{idk}(s)\left(K^L_{pk} s + K^L_{pik}\right) \end{aligned}.$$

The damping of dominant subsynchronous oscillation mode is calculated from (16) and (19) as

$$d^L_s = \frac{V_{k,0}{}^2 K^L_{pk} - \left(N I_{k,0} X_L\right)^2 \left(K^L_{pik} p_{Ik}/\omega_s + K^L_{pk} p_{Rk}\right)}{2 V_{k,0} C_k V_{dck,0} - \left(N I_{k,0} X_L\right)^2 K^L_{pk} p_{Ik}/\omega_s}, \tag{20}$$

where fast dynamics of ICC loop and PLL are represented by $G^L_{idk}(s) \approx 1$ and $p_{\theta k}(j\omega_s) = p_{Rk} + jp_{Ik} \approx 1$ [31], where $\omega_s$ is the angular frequency of subsynchronous oscillation.

From (20), the oscillation damping changes significantly as the number of ESSs, i.e., $N$, increases, which is defined as the dynamic superimposition effect of GFL-CESSs. The dynamic superimposition is equivalent to amplifying the output AC current by a factor of $N$ times. This implies the aggregated effect of such GFL-CESSs is identical to a GFL-ESS with $N$-times AC current. Similar conclusions align with the capacity equivalence principle observed in large-scale integration of renewable energy. It can be seen from (20) that a larger number of ESSs will induce stronger dynamic superimposition of GFL-CESSs, which reduces the oscillation damping and negatively impacts stability.

Similarly to (18), the transfer function of the power system with GFM-CESSs can be obtained from (12) to be

$$\mathbf{Y}^M_{xyk}(s) - \left(N\begin{bmatrix} 0 & -X_L \\ X_L & 0 \end{bmatrix}\right)^{-1} = \begin{bmatrix} 0 & 0 \\ 0 & 0 \end{bmatrix}, \tag{21}$$

where,

$$\mathbf{Y}^M_{xyk}(s) = \begin{bmatrix} K^M_{dk}(s) & 0 \\ 2K_{ipk}(s) K^M_{qk}(s)\left(I_{k,0} - K^M_{qk}(s)\right) & 2K^M_{qk}(s) - I_{k,0} \end{bmatrix},$$

$$K^M_{dk}(s) = -\left(K^M_{pk} + K^M_{pik} s^{-1}\right),\ K^M_{qk}(s) = -\left(K^M_{qk} + K^M_{qik} s^{-1}\right).$$

The characteristic equation of GFM-CESSs can be obtained by calculating determinant of (21), i.e., |Eq (21)|, as

$$F_M(s) + F_{MO}(s) = 0, \tag{22}$$

where,

$$\begin{aligned} F_M(s) = {} & K^M_{pk}\left(2K^M_{qk} + I_{k,0}\right)s^2 + 2K^M_{pik} K^M_{qik} + \\ & \left(K^M_{pik}(2K^M_{qk} + I_{k,0}) + 2K^M_{qik} K^M_{pk}\right)s \end{aligned},$$

$$F_{MO}(s) = Y_L \begin{bmatrix} Y_L s^2 + 2K_{ipk} K^M_{qk}\left(I_{k,0} + K^M_{qk}\right)s + \\ 2K_{ipk} K^M_{qik}\left(I_{k,0} + 2K^M_{qk}\right) + 2\dfrac{K_{ipk} K^{M\,2}_{qik}}{s} \end{bmatrix},\ Y_L = X_L{}^{-1}.$$

The damping of dominant subsynchronous oscillation mode is calculated from (16) and (22) as

$$d^M_s = \frac{K^M_{pik}\left(2K^M_{qk} + I_{k,0}\right) + 2K^M_{qik} K^M_{pk} + d_O}{2K^M_{pk}\left(2K^M_{qk} + I_{k,0}\right)}, \tag{23}$$

where,

$$d_O = \frac{\mathrm{Im}\left[F_O(j\omega_s)\right]}{\omega_s} = 2\frac{Y_L}{N}\left(K_{ipk} K^M_{qk}\left(I_{k,0} + K^M_{qk}\right) - \frac{2K_{ipk} K^{M\,2}_{qik}}{\omega_s{}^2}\right).$$

In (23), the oscillation damping changes significantly as the ESS number increases, which is the dynamic superimposition of GFM-CESSs. However, the dynamic superimposition of GFM-CESSs is different from that of GFL-CESSs. It can be seen from (23) that the damping consists of two parts. The one is determined by control parameters and operational states (unrelated to ESS number). The other one strengthens as ESS number decreases.

### *C. Conditions of Positive Dynamic Superimposition of GFM-CESSs*

The dynamic superimposition of GFM-CESSs remains valid as long as ESSs are identical or similar. However, its influence on oscillation damping is not always positive. For GFM-CESSs, self-stability is inherently satisfied, meaning that $I_{k,0} + 2K_{qk}^M$ is always positive. Thus, the impact of dynamic superimposition on oscillation damping is positive when $d_O > 0$. Under this condition, the following condition can be derived from (23) as

$$\omega_s^{\ 2} K_{qk}^M \left( I_{k,0} + K_{qk}^M \right) - 2K_{qik}^{M\ 2} > 0 . \tag{24}$$

From (24), it can be concluded that, unlike GFL-CESSs, the dynamic superimposition of GFM-CESSs enhances oscillation damping and improves stability when the condition in (24) is satisfied (which ensures $d_O > 0$). Based on (24), the parameter region of $K_{qk}^M$ and $2K_{qik}^M$ that ensures stability is illustrated in Fig. 5 for different oscillation frequencies.

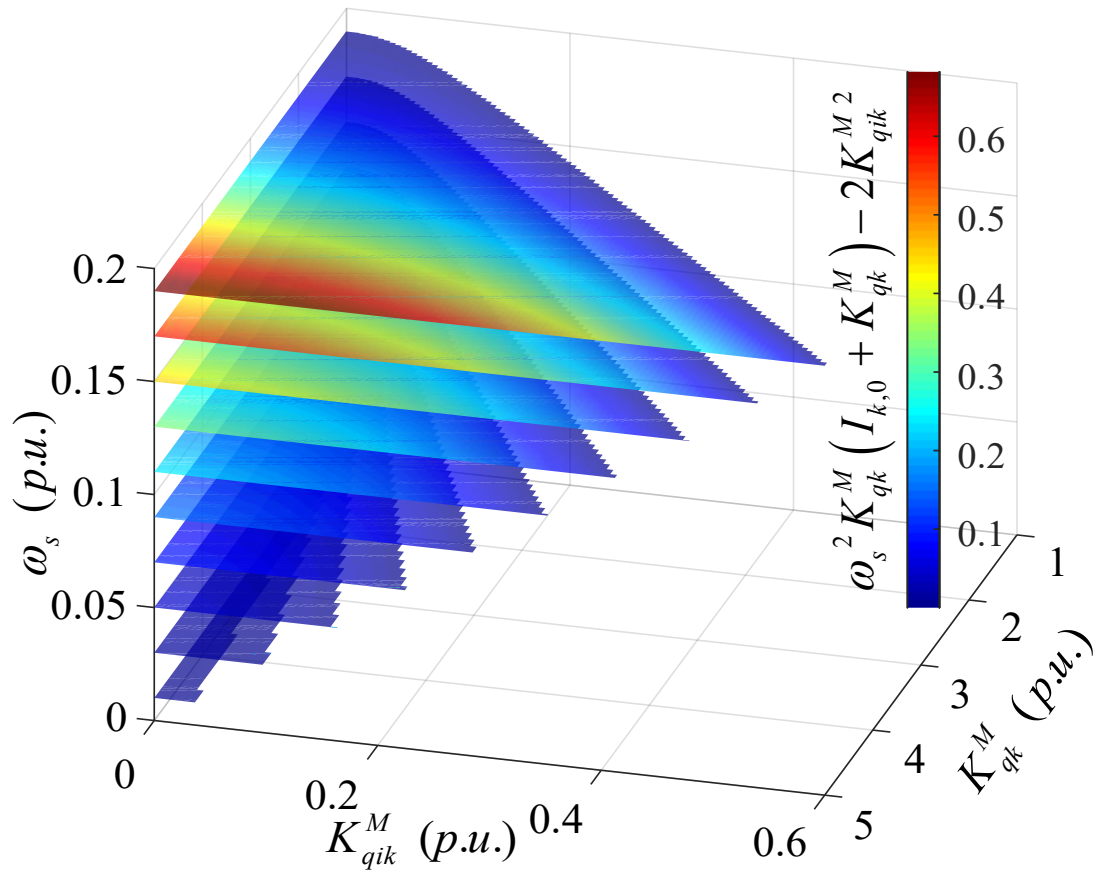

Fig. 5. Parameter region where dynamic superimposition improves damping.

In Fig. 5, $I_{k,0}$ = -1 p.u. is selected because it corresponds to the operating point with the worst damping. It can be observed that the parameter region varies under different frequencies, and expands significantly at higher frequency. Furthermore, a larger value of $K_{qk}^M$ is more beneficial for enhancing system stability.

### *D. Comparative Damping Mechanism Analysis of GFL-CESSs and GFM-CESSs with Respect to Bidirectional Power Flow*

To separate the current-related terms, the expression of (20) can also be reformulated as

$$d_s^L = \kappa_1 - \kappa_2 I_{k,0}^{\ 2} , \tag{25}$$

where $\kappa_1$ and $\kappa_2$ are constants that are independent of $I_{k,0}$. They can be expressed as

$$\begin{aligned} \kappa_1 &= \frac{V_{k,0}^{\ 2} K_{pk}^L}{2V_{k,0} C_k V_{dck,0} - \left( N I_{k,0} X_L \right)^2 K_{pk}^L p_{Ik} / \omega_s} \\ \kappa_2 &= \frac{\left( N X_L \right)^2 \left( K_{pik}^L p_{Ik} / \omega_s + K_{pk}^L p_{Rk} \right)}{2V_{k,0} C_k V_{dck,0} - \left( N I_{k,0} X_L \right)^2 K_{pk}^L p_{Ik} / \omega_s} \end{aligned} .$$

From (25), it is obvious that the oscillation damping is dominated by the amplitude and is insensitive to the direction of the current. Therefore, increasing the output current reduces the damping for both positive ($I_{k,0} > 0$) and negative ($I_{k,0} < 0$) current, indicating that, in GFL-CESSs, the movement of the oscillation damping is weakly dependent on the power flow direction.

The derivative of oscillation damping with respect to output current under GFM control is calculated from (23) as

$$\Delta d_s^M = \kappa_3 \Delta I_k , \tag{26}$$

where,

$$\kappa_3 = \frac{2Y_L K_{ipk} K_{qk}^M + N K_{pik}^M - 2N d_s^M K_{pk}^M}{2N K_{pk}^M \left( 2K_{qk}^M + I_{k,0} \right)} > 0 .$$

$\kappa_3$ is positive because $I_{k,0} + 2K_{qk}^M$ is always positive since the self-stability of GFM control must be satisfied.

Comparing (25) and (26), the impact of power flow on the oscillation damping of GFM-CESSs differs from that of GFL-CESSs. Specifically, the oscillation damping improves when the output current increases in the positive direction ($I_{k,0} > 0$ and $\Delta I_{k,0} > 0$) and decreases when the output current increases in the negative direction ($I_{k,0} < 0$ and $\Delta I_{k,0} < 0$). This indicates that the direction of oscillation damping movement is significantly affected by the power flow direction for GFM-CESSs.

The impact of the amplitude variation of power flow on the oscillation damping is also different between GFL-CESSs and GFM-CESSs.

The derivative of the oscillation damping with respect to the output current under GFL control is obtained from (25) as

$$\Delta d_s^L = \kappa_4 \Delta I_k = -2\kappa_2 I_{k,0} \Delta I_k . \tag{27}$$

It can be seen from (26) and (27) that the maximum values of $\Delta d_s^M$ and $\Delta d_s^L$ occur at $I_{k,0}$ = -1 p.u. Therefore, we can derive that

$$\begin{aligned} \max\left( \left| \kappa_3 \right| \right) &> \frac{2Y_L K_{ipk} K_{qk}^M + N K_{pik}^M - 2N d_s^M K_{pk}^M}{4N K_{pk}^M K_{qk}^M} \\ &= \frac{1}{N X_L} \frac{K_{ipk}}{2K_{pk}^M} + \frac{K_{pik}^M}{4K_{pk}^M K_{qk}^M} - \frac{d_s^M}{2K_{qk}^M} \\ \max\left( \left| \kappa_4 \right| \right) &\le \frac{2\left( N X_L \right)^2 \left( K_{pik}^L p_{Ik} / \omega_s + K_{pk}^L p_{Rk} \right)}{2V_{k,0} C_k V_{dck,0} - \left( N I_{k,0} X_L \right)^2 K_{pk}^L p_{Ik} / \omega_s} \\ &\approx \left( N X_L \right)^2 \frac{K_{pk}^L}{C_k} \end{aligned} , \tag{28}$$

where, the condition $p_{\theta k}(j\omega_s) \approx 1$ is used, and the AC and DC voltages are simplified as $V_{k,0}$ = 1 p.u. and $V_{dck,0}$ = 1 p.u.

Considering that $NX_L$ is limited to be smaller than 1 p.u., we have $1/(NX_L) >> (NX_L)^2$. Thus, (28) changes to

$$\left| \kappa_3 \right| > \frac{1}{N X_L} \frac{K_{ipk}}{2K_{pk}^M} + \frac{K_{pik}^M}{4K_{pk}^M K_{qk}^M} - \frac{d_s^M}{2K_{qk}^M} >> \left( N X_L \right)^2 \frac{K_{ipk}}{2K_{pk}^M} . \tag{29}$$

Therefore, the sensitivity of the oscillation damping with respect to the power amplitude is more significant for GFM-CESSs than for GFL-CESSs, as long as $0.5K_{ipk}/K_{pk}$ is of the same order or higher than $K_{pk}/C_k$. This conclusion is also physically reasonable: GFM-CESSs provide both active and reactive power support to the connected power system, thereby enhancing the coupling between the GFM-CESSs and the system and leading to a more significant effect caused by power flow. In contrast, GFL-CESSs are designed to follow the power system as quickly as possible, and thus the dynamic coupling is normally weak. Therefore, the impact from power-flow amplitude of GFL-CESSs is comparatively smaller.

### *E. Comparative Damping Mechanism Analysis of GFL-CESSs and GFM-CESSs with Respect to Angle-related Control Loops*

The derivative of oscillation damping with respect to angle-related control loop parameters (PLL of GFL control and PSC loop of GFM control) is obtained in (30) and (33), respectively.

The derivative of oscillation damping with respect to PLL parameters ($K^L_{\theta pk}$ and $K^L_{\theta ik}$) is calculated from (20) as

$$\begin{aligned}\frac{\Delta d_s^L}{\Delta K_{\theta ik}^L} &= T_{dpk}\frac{\Delta p_{\theta k}(s)}{\Delta K_{\theta ik}^L} = -\frac{T_{dpk}V_{k,0}s^2}{\left(s^2 - V_k\left(K_{\theta pk}^L s + K_{\theta ik}^L\right)\right)^2}\\ \frac{\Delta d_s^L}{\Delta K_{\theta pk}^L} &= T_{dpk}\frac{\Delta p_{\theta k}(s)}{\Delta K_{\theta pk}^L} = -\frac{T_{dpk}V_{k,0}s^3}{\left(s^2 - V_{k,0}\left(K_{\theta pk}^L s + K_{\theta ik}^L\right)\right)^2}\end{aligned}, \quad (30)$$

where the item in denominator represents the inherent dynamic of PLL. If an open-loop pole of which is denoted as $\lambda_p = -d_p + j\omega_p$, the denominator in (30) can be rewritten as

$$s^2 - V_k\left(K_{\theta pk}^L s + K_{\theta ik}^L\right) = \left(s-\lambda_p\right)\left(s-\lambda_p^*\right), \quad (31)$$

where * represents conjugation of an eigenvalue.

Considering that the inherent frequency of the PLL is much higher than that of the outer control loop for rapidly locking the voltage angle, the difference between $\omega_p$ and $\omega_s$ is large (i.e., $\omega_p >> \omega_s$) in the case studied in this paper, which yields

$$\begin{aligned}\left.\frac{\Delta d_s^L}{\Delta K_{\theta ik}^L}\right|_{s=j\omega_s} &= T_{dpk}\frac{V_{k,0}\omega_s^2}{\left(\left(j\omega_s-\lambda_p\right)\left(j\omega_s-\lambda_p^*\right)\right)^2} \approx 0\\ \left.\frac{\Delta d_s^L}{\Delta K_{\theta pk}^L}\right|_{s=j\omega_s} &= T_{dpk}\frac{jV_{k,0}\omega_s^3}{\left(\left(j\omega_s-\lambda_p\right)\left(j\omega_s-\lambda_p^*\right)\right)^2} \approx 0\end{aligned}. \quad (32)$$

The results in (32) demonstrate that the impact of PLL on the focused stability in this paper is weak if the parameters of PLL have been well designed. Therefore, dynamics of PLL in the low-frequency range can be considered as $p_{\theta k}(j\omega_s) = p_{Rk} + jp_{Ik} \approx 1$ in (20).

The derivative of oscillation damping with respect to PSC loop parameter ($K^M_{ipk}$) is calculated from (23) as

$$\frac{\Delta d_s^M}{\Delta K_{ipk}} = \frac{Y_L}{N}\frac{K_{qk}^M\left(I_{k,0}+K_{qk}^M\right)\omega_s^2 - 2K_{qik}^{M\,2}}{K_{pk}^M\left(2K_{qk}^M+I_{k,0}\right)\omega_s^2}. \quad (33)$$

Different from the conclusion drawn for GFL control, the impact of the PSC loop on subsynchronous oscillation cannot be ignored. This impact is amplified either the line admittance increases or the number of GFM-CESSs decreases.

### *F. Comparative Damping Mechanism Analysis of GFL-CESSs and GFM-CESSs with Respect to d-q Control Loops*

The derivative of oscillation damping with respect to *d*-axis control parameters is obtained in (34) for GFL control and in (35) for GFM control.

$$\frac{\Delta d_s^L}{\Delta K_{pk}^L} = \frac{V_{k,0}^2 - \left(NI_{k,0}X_L\right)^2}{2V_{k,0}C_kV_{dck,0}}, \frac{\Delta d_s^L}{\Delta K_{pik}^L} \approx 0. \quad (34)$$

$$\frac{\Delta d_s^M}{\Delta K_{pk}^M} = -\frac{K_{pik}^M\left(2K_{qk}^M+I_{k,0}\right)+d_O}{\left(2K_{pk}^M\left(2K_{qk}^M+I_{k,0}\right)\right)^2}, \frac{\Delta d_s^M}{\Delta K_{pik}^M} = \frac{1}{2K_{pk}^M}. \quad (35)$$

Comparing (34) with (35), the impact of *d*-axis control loop parameters of GFL-CESSs on stability is primarily affected by $K^L_{pk}$ and rarely related to $K^L_{pik}$, indicating that the impact of these parameters is approximately decoupled. However, for GFM-CESSs, the damping is affected by both parameters, with the significant difference occurring in the $K^M_{pik}$ and $K^L_{pik}$. Therefore, the impact of *d*-axis control loop parameters involves stronger couplings when GFM control is adopted, which should be carefully considered.

With conventional vector control, decoupling between the *d*- and *q*-axis dynamics is well established. Accordingly, when the dynamics are dominated by the *d*-axis control loops, the *q*-axis control loop parameters of GFL-CESSs do not affect the dynamics. This is also verified by (36), in which the derivative of the oscillation damping with respect to the *q*-axis control loop parameters under GFL control equals zero.

$$\frac{\Delta d_s^L}{\Delta K_{qk}^L} = 0, \frac{\Delta d_s^L}{\Delta K_{qik}^L} = 0. \quad (36)$$

The derivative of oscillation damping with respect to *q*-axis control loop parameters under GFM control is

$$\begin{aligned}\frac{\Delta d_s^M}{\Delta K_{qk}^M} &= \frac{K_{pik}^M/K_{pk}^M - 2d_s^M}{2K_{qk}^M+I_{k,0}} + \frac{Y_L}{N}\frac{K_{ipk}}{K_{pk}^M}\\ \frac{\Delta d_s^M}{\Delta K_{qik}^M} &= \frac{1}{2K_{qk}^M+I_{k,0}} - \frac{Y_L}{N}\frac{4K_{ipk}K_{qik}^M}{K_{pk}^M\left(2K_{qk}^M+I_{k,0}\right)\omega_s^2}\end{aligned}. \quad (37)$$

In (37), the impact of the *q*-axis control-loop parameters on the oscillation damping is significant, comparable to that of the *d*-axis control loop. It indicates that dynamic coupling exists between the *d*- and *q*-axis control loops and is non-negligible. Therefore, GFL-CESSs exhibit weak dynamic coupling between the *d-q* control loops, whereas the dynamic coupling in GFM-CESSs is strong, which complicates stability analysis and risk factor identification for GFM-CESSs.

TABLE I
Selection of different control-based CESSs for typical scenarios

| Scenarios | Recommendations |
|---|---|
| Strong grid connections | Dynamic coupling of GFM control intensifies under strong grid connections, adversely affecting stability. *GFL-CESSs are recommended. Seeing [5] for details.* |
| Weak grid connections | PLL instability risks dominate under weak grid conditions, whereas GFM control enhances stability. *GFM-CESSs are recommended. Seeing [17] for details.* |
| Large-scale integration | Increasing GFM-ESS number improves oscillation damping (opposite for GFL-CESSs). *GFM-CESSs are recommended. Seeing Section III-B and C. Verifications are in Section VI-B.* |
| Bidirectional active power | Oscillation modes exhibit larger deviations and higher sensitivity to operational states with GFM control. *GFL-CESSs are recommended. Seeing Section III-D. Verifications are in Section VI-A.* |
| Requiring weak control loop coupling | Multiple control loops of GFM control are coupled and jointly affect the stability, whereas *d*- and *q*-axis control loops of GFL control remain decoupled. *GFL-CESSs are recommended. Seeing Section III-E and F. Verifications are in Section VI-C and D.* |

The compared results are summarized in Table I. It can be observed that a single-type control cannot be well applied across diverse scenarios. Specifically, GFM-CESSs demonstrate better performance under large-scale integration and weak grid connections, whereas GFL-CESSs are preferred for strong grid connections and scenarios involving significant bidirectional active power and requiring weak dynamic coupling within control loops. To better match diverse scenarios, hybrids of GFL and GFM controls are usually adopted. However, the ratio of GFM-CESSs in hybrid CESSs must be carefully constrained to maintain stability margins.

## IV. Damping Analysis of Dominant Oscillation Mode in CESSs with Hybrid Controls

### A. Characteristic Equation under Hybrid Controls

The transfer function of the power system connected with hybrid CESSs, including $N_1$ GFL-CESSs and $N_2$ GFM-CESSs, can be derived from (18) and (21) as

$$\mathbf{H}_H(s) = N_1\mathbf{Y}_{xyk}^L(s) + N_2\mathbf{Y}_{xyk}^M(s) - \left(\begin{bmatrix} 0 & -X_L \\ X_L & 0 \end{bmatrix}\right)^{-1} = \mathbf{0}\,. \quad (38)$$

$\mathbf{H}_H(s)$ can be divided into two parts: one corresponding to GFL-CESSs ($\mathbf{H}_L(s)$) and the other to GFM-CESSs ($\mathbf{H}_M(s)$):

$$\mathbf{H}_H(s) = \mathbf{H}_L(s) + \mathbf{H}_M(s)\,, \quad (39)$$

where $N_1 + N_2 = N$,

$$\mathbf{H}_L(s) = N_1\left(\mathbf{Y}_{xyk}^L(s) - \left(N\begin{bmatrix} 0 & -X_L \\ X_L & 0 \end{bmatrix}\right)^{-1}\right),$$

$$\mathbf{H}_M(s) = N_2\left(\mathbf{Y}_{xyk}^M(s) - \left(N\begin{bmatrix} 0 & -X_L \\ X_L & 0 \end{bmatrix}\right)^{-1}\right).$$

The dominant oscillation mode of (39) can be calculated by

$$|\mathbf{H}_H(s)| = |\mathbf{H}_L(s)| + |\mathbf{H}_M(s)| + h_O(s) = 0\,, \quad (40)$$

where $h_O(s) = \mathrm{Tr}(\mathbf{H}_L(s))\mathrm{Tr}(\mathbf{H}_M(s)) + \mathrm{Tr}(\mathbf{H}_L(s)\mathbf{H}_M(s))$, Tr() represents the trace of a square matrix.

Denote the dominant oscillation modes of $\mathbf{H}_L(s)$ and $\mathbf{H}_M(s)$ as $\lambda_s^L$ and $\lambda_s^M$, respectively. The characteristic equation of (40) can be rewritten as

$$\overbrace{\frac{h_L(s)}{(s-\lambda_s^L)(s-\lambda_s^{L*})}}^{|\mathbf{H}_L(s)|} + \overbrace{(s-\lambda_s^M)(s-\lambda_s^{M*})h_M(s)}^{|\mathbf{H}_M(s)|} = -h_O(s)\,. \quad (41)$$

### B. Strong Dynamic Interaction between GFL-CESSs and GFM-CESSs

For a concerned oscillation frequency $\omega_s$, and a second-order oscillation loop related to $\lambda_s^L$, the transfer function of this loop and the remaining power system (excluding it) can be expressed in the form of (14) from (41) as

$$T_O(s) + T_R(s) = 0\,, \quad (42)$$

where,

$$T_O(s) = (s-\lambda_s^L)(s-\lambda_s^{L*})$$

$$T_R(s) = \frac{h_L(s)}{h_O(s) + (s-\lambda_s^M)(s-\lambda_s^{M*})h_M(s)}\,.$$

The variation in the damping coefficient of the second-order oscillation loop, as affected by the remaining power system at $\omega_s$, can be calculated according to (16) as

$$\Delta b = \frac{\mathrm{Im}[T_R(j\omega_s)]}{\omega_s} = \frac{1}{\omega_s}\mathrm{Im}\left[\frac{h_L(j\omega_s)}{h_O(j\omega_s) + (j\omega_s-\lambda_s^M)(j\omega_s-\lambda_s^{M*})h_M(j\omega_s)}\right]. \quad (43)$$

From (43), it can be verified that the maximum value of $\Delta b$ occurs when the oscillation frequency of $\lambda_s^M$ is equal to $\omega_s$, Under this condition, $\lambda_s^L$ will undergo a significant change and may become unstable if $\Delta b < 0$.

Similarly, if the second-order oscillation loop related to $\lambda_s^M$ is considered, the transfer function of this loop and the remaining power system (excluding it) can be expressed as

$$T_O(s) = (s-\lambda_s^M)(s-\lambda_s^{M*})$$
$$T_R(s) = \frac{h_O(s)}{h_M(s)} + \frac{h_L(s)}{h_M(s)(s-\lambda_s^L)(s-\lambda_s^{L*})}\,. \quad (44)$$

The variation in the damping coefficient of this second-order oscillation loop, as affected by the remaining power system at $\omega_s$, can be calculated following (42) to (43) as:

$$\Delta b = \frac{\mathrm{Im}[T_R(j\omega_s)]}{\omega_s} = \frac{1}{\omega_s}\mathrm{Im}\left[\frac{h_O(j\omega_s)}{h_M(j\omega_s)} + \frac{h_L(j\omega_s)}{h_M(j\omega_s)(j\omega_s-\lambda_s^L)(j\omega_s-\lambda_s^{L*})}\right]. \quad (45)$$

From (45), it can be verified that the maximum value of $\Delta b$ occurs when the oscillation frequency of $\lambda_s^L$ is equal to $\omega_s$. Under this condition, $\lambda_s^M$ will undergo a significant change and may become unstable if $\Delta b < 0$.

Therefore, by combining the results of (43) and (45), it is evident that if $\lambda_s^M$ is close to $\lambda_s^L$, the GFM-CESSs and GFL-CESSs will provide significant damping to each other's dominant oscillation loop, thereby substantially affecting the closed-loop oscillation modes. This interaction strongly affects oscillation damping and potentially leads to instability. This phenomenon is also commonly referred to as modal resonance [33]-[34].

Under initial conditions, the oscillation modes of GFL-CESSs and GFM-CESSs are designed to be well separated within stability limits. However, as demonstrated in Section III-C, significant bidirectional power flow causes the dominant oscillation mode of GFM-CESSs to change considerably. This effect is further amplified through dynamic aggregation with an increasing number of GFM-CESSs. Consequently, as their proportion grows, the dominant oscillation mode of GFM-CESSs shifts over a wider range, potentially approaching that of GFL-CESSs. Such proximity can lead to strong dynamic interactions between GFL-CESSs and GFM-CESSs, which in turn reduces the overall oscillation damping.

## V. Error Analysis of Modeling Simplifications

### A. Error Analysis for Simplification of Voltage Angle

In the derivation of (18), the voltage angle is assumed to be zero so that $\sin(\theta_{k,0}) = 0$, and consequently the off-diagonal elements are zero. However, in practice $\theta_{k,0}$ may be close to zero but not exactly zero. For the diagonal elements, the impact of considering the voltage angle can be treated as a variation in their original parameters, since the transfer functions related to $\theta_{k,0}$ are all linearized. This does not change the order of $H_{ddk}(s)$ and $p_{\theta k}(s)$. Therefore, we focus on how the off-diagonal elements of $\mathbf{Y}^L_{xyk}(s)$ affect the analyzed results. For generality, $\mathbf{Y}^L_{xyk}(s)$ is written as

$$\mathbf{Y}^L_{xyk}(s) = \begin{bmatrix} H_{ddk}(s) & e_{\theta 1}(s) \\ e_{\theta 2}(s) & p_{\theta k}(s) I_{k,0} / V_{k,0} \end{bmatrix}. \tag{46}$$

Replacing $\mathbf{Y}^L_{xyk}(s)$ in (18) by that in (46), the characteristic equation of GFL-CESSs can be obtained by calculating |(18)| as

$$\begin{aligned} &N^2 X_L{}^2 I_{k,0} p_{\theta k}(s) H_{ddk}(s) + V_{k,0} \\ &+ e_{\theta 1} e_{\theta 2} N^2 X_L{}^2 V_{k,0} + N X_L (e_{\theta 1} + e_{\theta 2}) V_{k,0} = 0 \end{aligned}. \tag{47}$$

The off-diagonal elements of $\mathbf{Y}^L_{xyk}(s)$ should be written as voltage angle-related transfer functions, which equal zero when $\theta_{k,0} = 0$. Moreover, considering that $\sin\theta_{k,0}$ is significantly less than 1, i.e., $\sin\theta_{k,0} << 1$, and the frequency range of interest in this paper is also much less than 1 p.u., i.e., $|s| < 1$ with $s = j\omega_s$, the higher-order infinitesimal coefficients ($e_{\theta 1} e_{\theta 2}$ related items) in (47) can be neglected, and the form of $e_{\theta 1} + e_{\theta 2}$ can be written as

$$e_{\theta 1} + e_{\theta 2} = e_{\theta k} \sin\theta_{k,0}. \tag{48}$$

Substituting (48) into (47), $F_L(s)$ in (19) changes to

$$F_L(s) = \left(V_{k,0} + e_{\theta k} \sin\theta_{k,0}\right) C_k V_{dck,0} s^2 + V_{k,0} G^L_{idk}(s) \left(K^L_{pk} s + K^L_{pik}\right). \tag{49}$$

Therefore, it can be seen that the impact of the voltage angle is equivalent to a variation in the control and operating parameters of $F_L(s)$ caused by $e_{\theta k}\sin\theta_{k,0}$, which can refer to the analyses in Sections III-D, E, and F. However, the impact of the voltage angle is weak since $e_{\theta k}\sin\theta_{k,0}$ is much less than 1 p.u., and thus $V_{k,0} + e_{\theta k}\sin\theta_{k,0}$ is approximately equal to $V_{k,0}$.

### B. Error Analysis for ESS Inconsistencies

For CESSs, the control parameters remain the same to avoid the inconsistency of the ESSs. However, as time goes by, the operation states, such as the current output from each ESS, may be different even though the control parameters are the same. By considering the difference in the currents, the transfer function of the power system connected with GFL-CESSs can be obtained from (18) to be

$$\sum_{k=1}^{N} \left(\gamma_k \mathbf{Y}^L_{xyk}(s)\right) - \left(\begin{bmatrix} 0 & -X_L \\ X_L & 0 \end{bmatrix}\right)^{-1} = \begin{bmatrix} 0 & 0 \\ 0 & 0 \end{bmatrix}, \tag{50}$$

where $\gamma_k$ indicates the difference of the actual current of the $k^{\text{th}}$ ESS compared to the normal current, and $\gamma_k = 1$ if the ESSs are the same.

Equation (50) can be rewritten as

$$\bar{\gamma} \mathbf{Y}^L_{xyk}(s) - \left(N \begin{bmatrix} 0 & -X_L \\ X_L & 0 \end{bmatrix}\right)^{-1} = \begin{bmatrix} 0 & 0 \\ 0 & 0 \end{bmatrix}, \tag{51}$$

where $\bar{\gamma} = \frac{1}{N} \sum_{k=1}^{N} \gamma_k$.

The (51) indicates a conclusion that if the ESSs are different in current, the impact of CESSs on the oscillation damping is equal to that of CESSs with a current of $\gamma_k I_{k,0}$. And thus, the conclusions drawn in Section III-B, C, and D are applicable.

## VI. Case Study

The configuration of the examined case is shown in Fig. 1, and both GFM-CESSs and GFL-CESSs are involved. In the cases, a mathematical model and an electromagnetic transient (EMT) simulation model are used. The mathematical model is established in Section II, while the EMT model is implemented in Simulink. The external power system is represented by an infinite bus, considering that the bulk power system is GW-level, significantly larger than that of CESSs. The inductance of inner lines within CESSs is the same, 0.005 p.u., and is also much less than that of the transmission line. The CESSs are used for power support for the external power system. Therefore, there are no loads or any other sources in the region, and all power is exchanged between the CESSs and the external power system. Some parameters are listed in Table II, and the others are given following by specific cases, where only the parameters of the 1st ESS are given since those of all ESSs are the same. It should be noted that in figures of oscillation mode trajectory, only the dominant oscillation mode is illustrated, since it reflects the least-damped dynamics that can potentially lead to system instability.

TABLE II
Simulation and control parameters used in case studies

| **System parameters** | **Values** |
|---|---|
| Rated AC voltage | 260 V |
| Rated Power | 100 kW |
| Rated DC voltage | 500 V |
| AC impedance base | 0.676 Ω (1 p.u.) |
| AC inductance base | 0.00179 H (1 p.u.) |
| DC impedance base | 2.5 Ω (1 p.u.) |
| DC capacitance base | 1062 uF (1 p.u.) |
| AC filter of gird-side converter | 0.02 p.u. |
| DC capacitance of gird-side converter | 5 p.u. |
| Current control parameters | 0.2 + 100 rad/s |
| PLL parameters (for GFL control) | 5 + 500 rad/s |

| Number of ESSs | $N = 5$ |
|---|---|
| Inductance of inner lines within CESSs | 0.005 p.u. |
| Converter topology | Three-level [35] |
| Simulation step time | 2e-6 s |
| Control sample time | 1e-4 s |

*A. Comparative Analysis of the Impact of Bidirectional Power Flow on Oscillation Damping*

The trajectory of the subsynchronous oscillation mode when active power of each ESS varies between -1 p.u. to 1 p.u. is shown in Fig. 6. Here, the control parameters are: GFL-CESSs: $K_{pl}^{L} = 0.2$, $K_{ql}^{L} = 0.1$, $K_{pil}^{L} = 60$ rad/s, $K_{qil}^{L} = 10$ rad/s. GFM-CESSs: $K_{p1}^{M} = 3.5$, $K_{q1}^{M} = 1.6$, $K_{pi1}^{M} = 70$ rad/s, $K_{qi1}^{M} = 70$ rad/s, $K_{ip1}^{M} = 120$ rad/s. $L_L = 0.02$ p.u. It clearly shows that the impact of variation in active power on the dominant oscillation mode is significantly different when different controls are used.

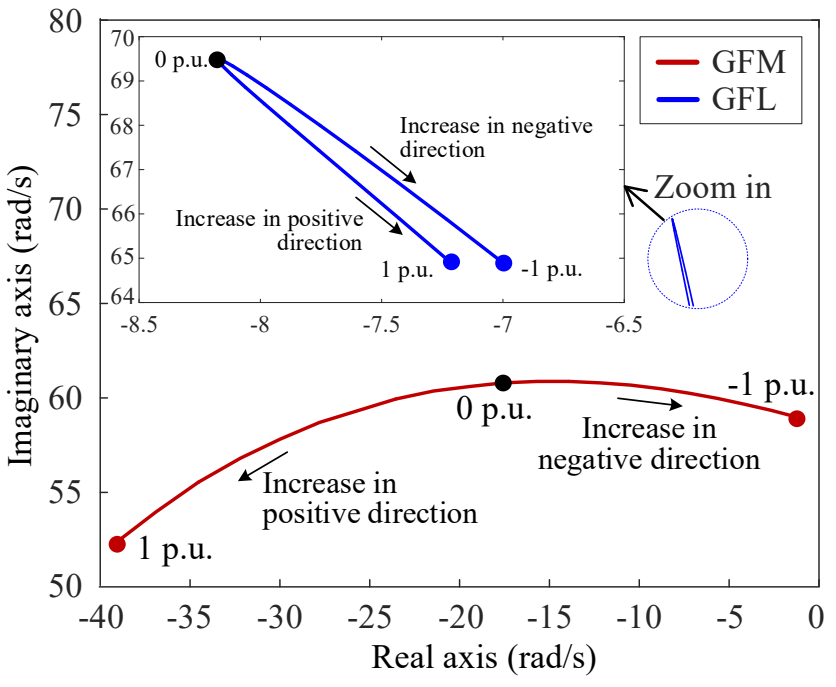

Fig. 6. Trajectories of oscillation modes when CESS active power varies under different controls.

It can be observed from Fig. 6 that the oscillation damping is more sensitive to power amplitude and direction for GFM-CESSs. In scenarios where power flow varies significantly, the oscillation mode may experience a significant variation, which requires attention. Specifically, the conclusions are organized from two perspectives, power amplitude and power direction, as follows.

Regarding the power amplitude, the impact of CESSs on dominant oscillation mode increases as the power amplitude rises for both GFL-CESSs and GFM-CESSs. However, such impact is more pronounced when GFM-CESSs are connected, with the oscillation damping varying from -17.57 rad/s to -39.3 rad/s as the power increases from 0 p.u. to 1 p.u. In contrast, under the same condition, the damping varies by only 1 rad/s for GFL-CESSs.

Regarding the power direction, the impact of CESS power direction on the dominant oscillation mode is different. The damping experiences opposite effects when the power direction reverses for GFM-CESSs, but the damping varies in a same direction for GFL-CESSs. For instance, the damping varies from -1.26 rad/s to -39.3 rad/s when power amplitude remains at 1 p.u. and direction reverses for GFM-CESSs. In contrast, under the same condition, the damping always moves to the right side for GFL-CESSs.

*B. Comparative Analysis of the Impact of Integration Number on Oscillation Damping*

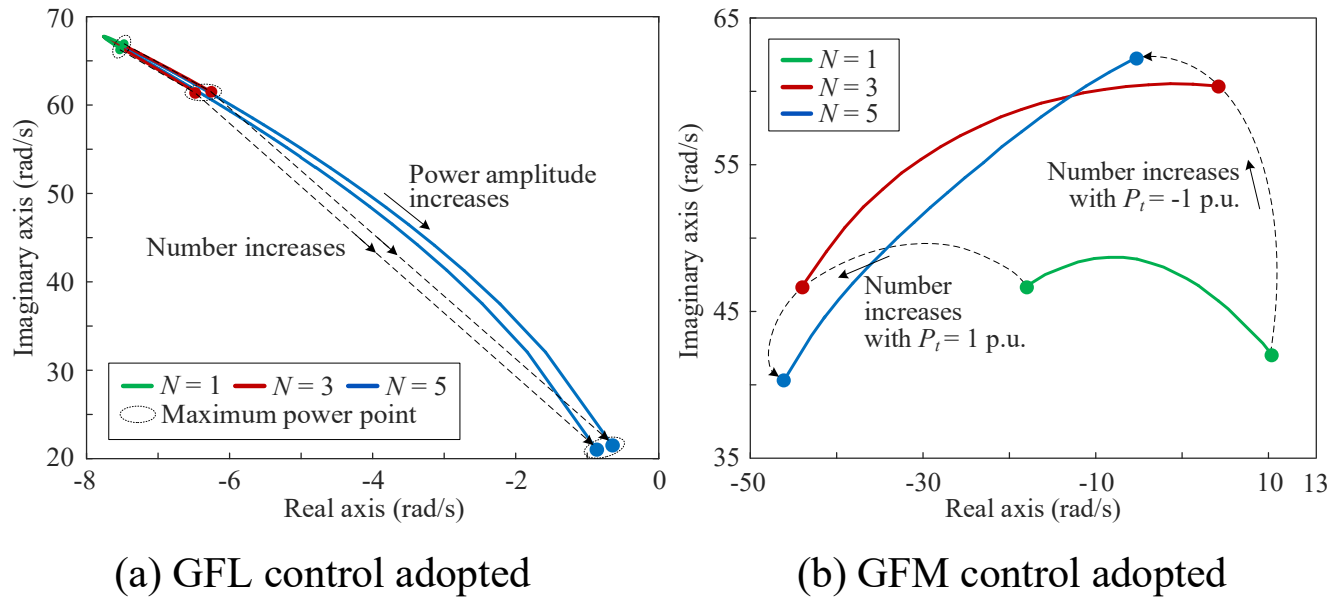

(a) GFL control adopted (b) GFM control adopted

Fig. 7. Trajectories of oscillation modes when number of ESSs increases.

In Fig. 7, the impact of increasing number of ESSs, $N$, on oscillation mode with different controls is illustrated. Here, the control parameters are: GFL-CESSs: $K_{pl}^{L} = 0.2$, $K_{ql}^{L} = 0.1$, $K_{pil}^{L} = 60$ rad/s, $K_{qil}^{L} = 10$ rad/s. GFM-CESSs: $K_{p1}^{M} = 3.0$, $K_{q1}^{M} = 1.0$, $K_{pi1}^{M} = 70$ rad/s, $K_{qi1}^{M} = 50$ rad/s, $K_{ip1}^{M} = 120$ rad/s. $L_L = 0.1$ p.u. The active power $P_t$ of each ESS varies from -1 p.u. to 1 p.u. It can be seen that regarding the trajectory with a certain number of ESSs, the movement trend of oscillation mode is similar to that shown in Fig. 6. That is, the oscillation mode of GFM-CESSs moves in a larger range than that of GFL-CESSs. However, from Fig. 7 (a), as the number of ESSs increases, the variation range of oscillation mode significantly increases, which is significant when the number of ESSs reaches five. The damping is close to being negative when the power amplitude reaches 1 p.u., indicating that increasing the ESS number of GFL-CESSs is risky of instability. However, the conclusions are significantly different when GFM control is used. It can be seen from Fig. 7 (b) that as the number of ESSs increases, the worst damping when each ESS operates at -1 p.u. improves from 10.2 rad/s to -5.5 rad/s, that is, GFM-CESSs vary from unstable to stable. Also, the stability at other operation points has also enhanced. It verifies the existence of dynamic superimposition in CESSs and highlights differences across CESSs with different controls.

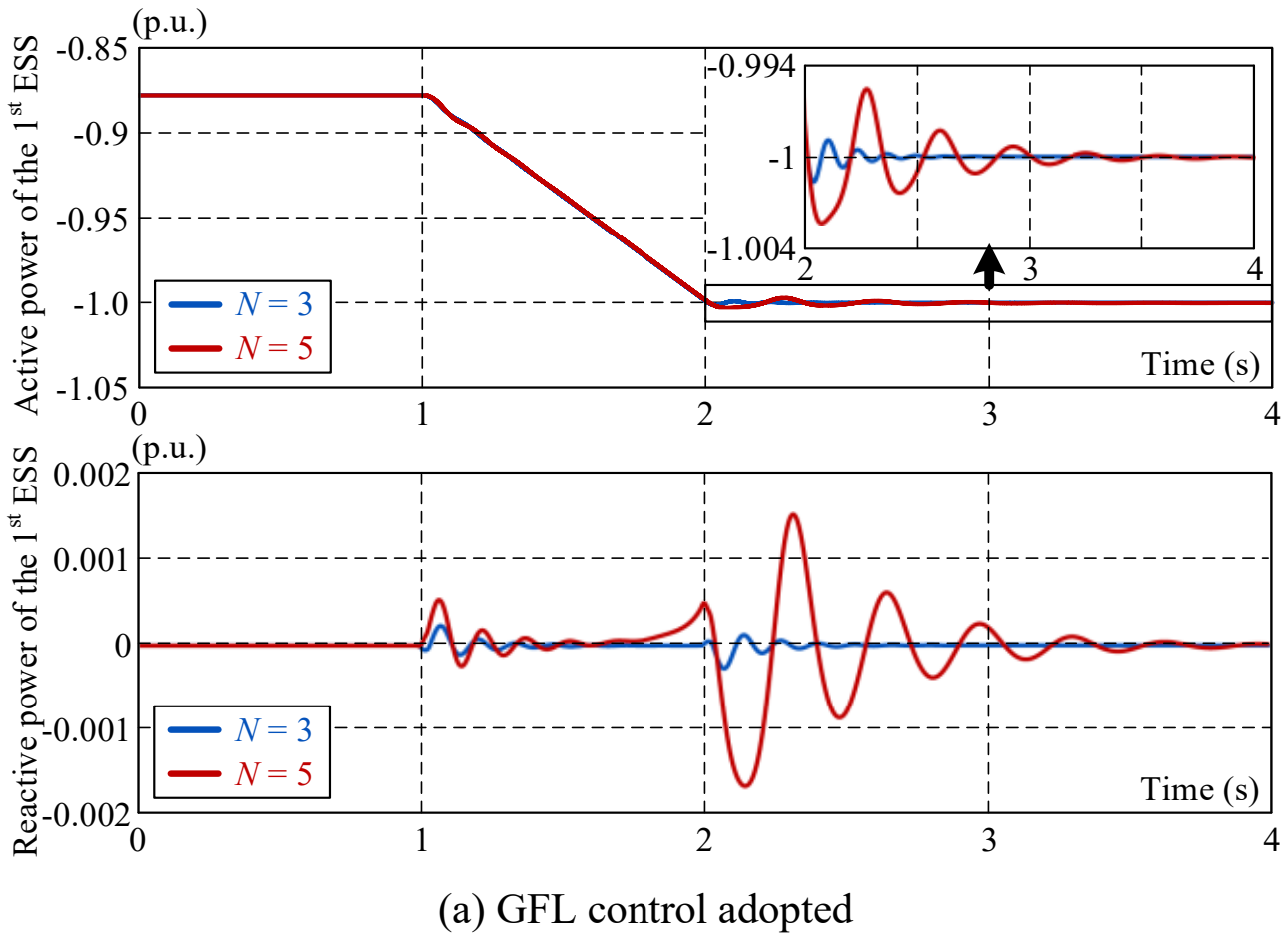

(a) GFL control adopted

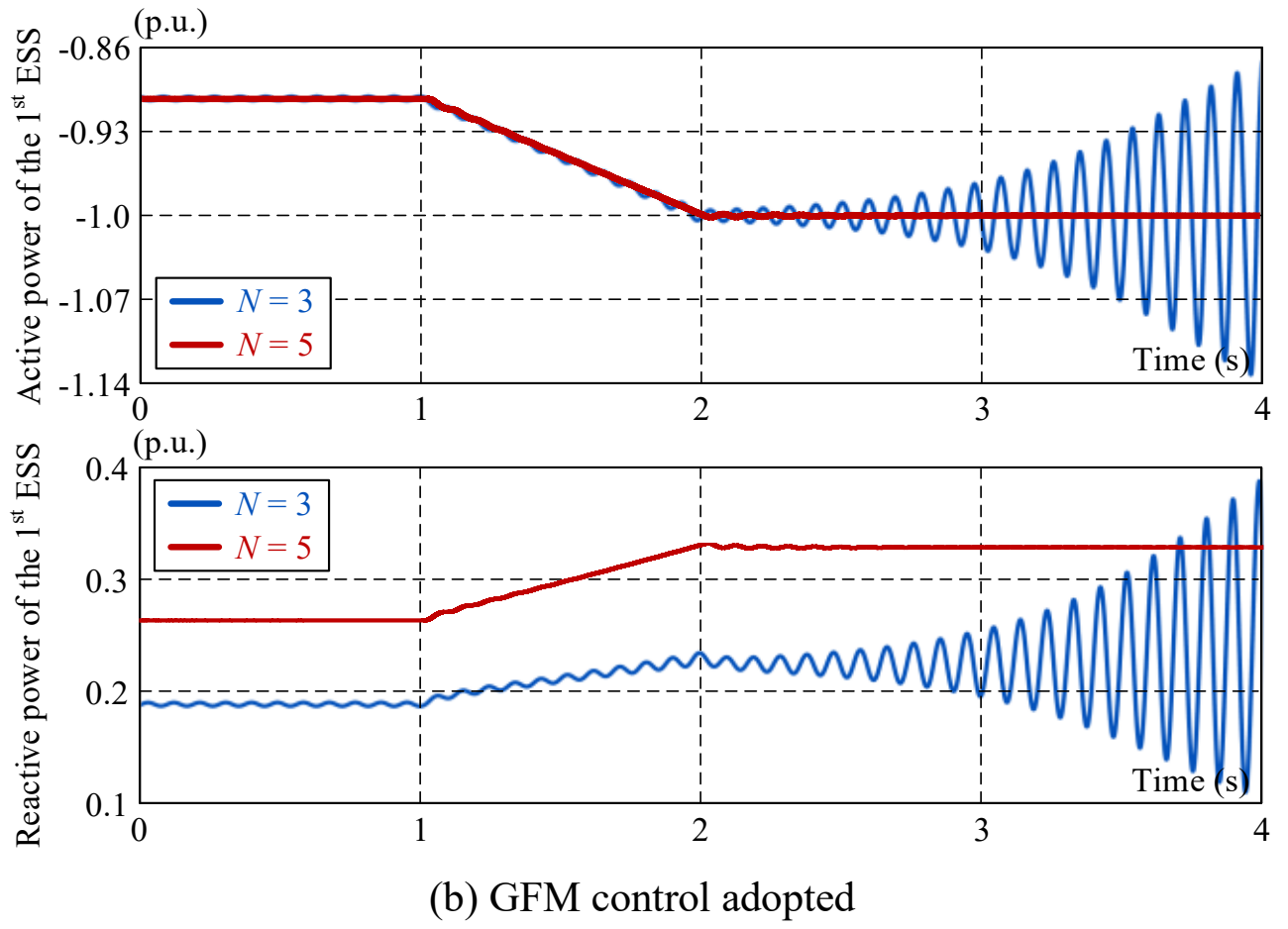

(b) GFM control adopted

Fig. 8. Time-domain simulation results when number of ESSs changes.

The time-domain simulation results in Simulink are shown in Fig. 8. During 1s to 2s, the reference of active power output from each ESS in GFL-CESSs changes from -0.88 p.u. to -1 p.u., and the reference of reactive power remains 0 p.u. The active and reactive power of GFL-CESSs are approximately decoupled, as oscillations in active power have only a minor influence on reactive power dynamics. For GFM-CESSs, the reference of active power output from each ESS changes from -0.9 p.u. to -1 p.u., and the reactive power suffers a significant change due to the coupled dynamics between active and reactive power. Furthermore, as ESS number of GFL-CESSs increases, oscillation damping decreases, whereas a larger ESS number of GFM-CESSs effectively mitigates oscillations.

Therefore, when considering large-scale CESSs, the worst damping may decrease, potentially leading to instability under GFL control. However, for large-scale GFM-CESSs, the worst damping can be improved. This suggests that for the GFM control discussed in this paper, no small-signal instability risk arises in the subsynchronous frequency range as the scale of GFM-CESSs increases. Thus, using GFM-CESSs is reasonable to improve the dynamics of CESSs.

### *C. Comparative Analysis of the Impact of Angle-related Control Loop Parameters on Oscillation Damping*

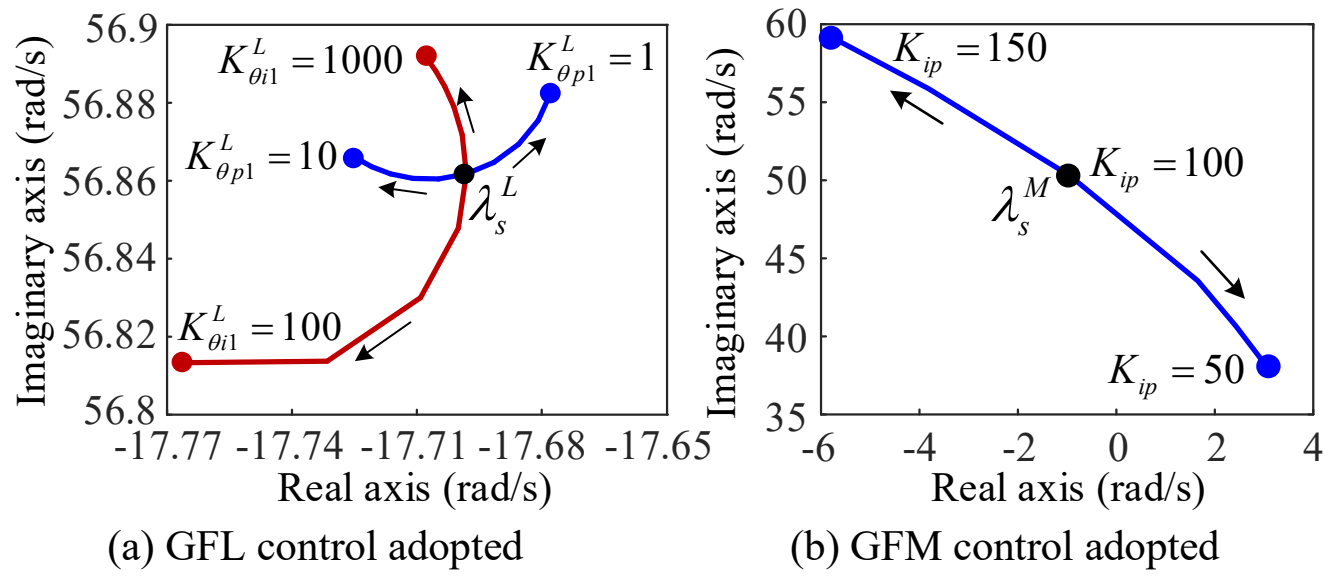

(a) GFL control adopted (b) GFM control adopted

Fig. 9. Trajectories of oscillation modes when angle-related parameters vary.

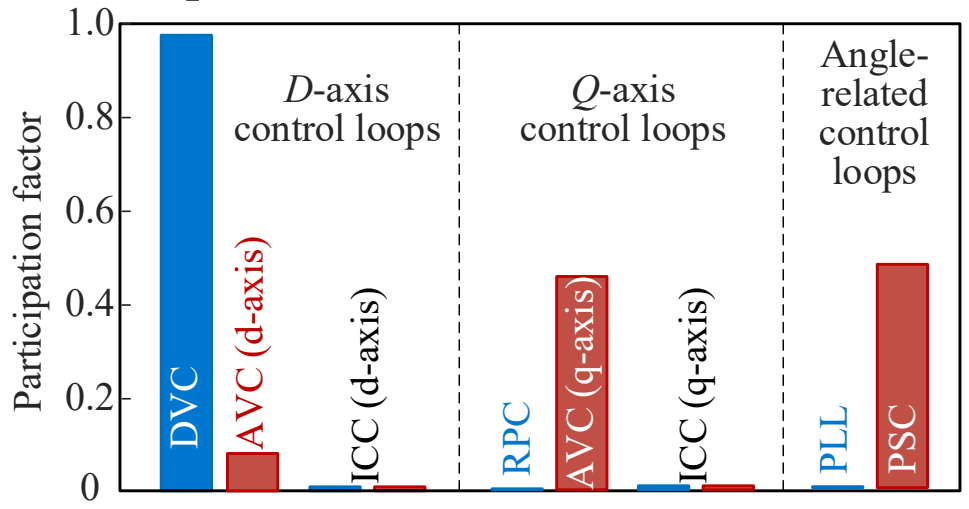

Fig. 10. Participation factors of oscillation modes when different controls used.

Subsections C and D adopt the same initial parameters. For GFL-CESSs, the initial parameters are set as $K_{p1}^L = K_{q1}^L = 0.5$, $K_{pi1}^L = K_{qi1}^L = 50$ rad/s, $L_L = 0.05$ p.u., $N = 5$. For GFM-CESSs, the initial parameters are set as $K_{p1}^M = 3$, $K_{q1}^M = 1.6$, $K_{pi1}^M = 80$ rad/s, $K_{qi1}^M = 50$ rad/s, $K_{ip1}^M = 100$ rad/s, $L_L = 0.02$ p.u., $N = 5$.

Examinations conducted for the PLL and PSC loop are illustrated in Fig. 9. The impact of angle-related control parameters of GFL-CESSs on the dominant oscillaiton mode is significantly weaker than those of GFM-CESSs. Therefore, the dynamics of various control loops of GFL-CESSs are normally decoupled, such that subsynchronous oscillation is only dominated by a single control loop. Differently, the control loops of GFM-CESSs are deeply coupled, and multiple control parameters can have an impact on the oscillation damping.

For further confirmation, the participation factor of dominant oscillation mode is calculated as shown in Fig. 10, where the results are highlighted in red for GFM-CESSs and in blue for GFL-CESSs. From the figure, it is clear that for GFL-CESSs, the dynamic characteristics are primarily influenced by the DC voltage control loop (i.e., the *d*-axis outer control loop). The participation of other control loops is negligible. Therefore, stability is mainly governed by the *d*-axis control loop when GFL control is applied. However, the results show significant differences for GFM-CESSs, where dynamic characteristics are simultaneously affected by the *d*-axis voltage control loop, *q*-axis voltage control loop, and angle-related control loop. Thus, the oscillation damping under GFM-CESSs is determined by multiple control loops.

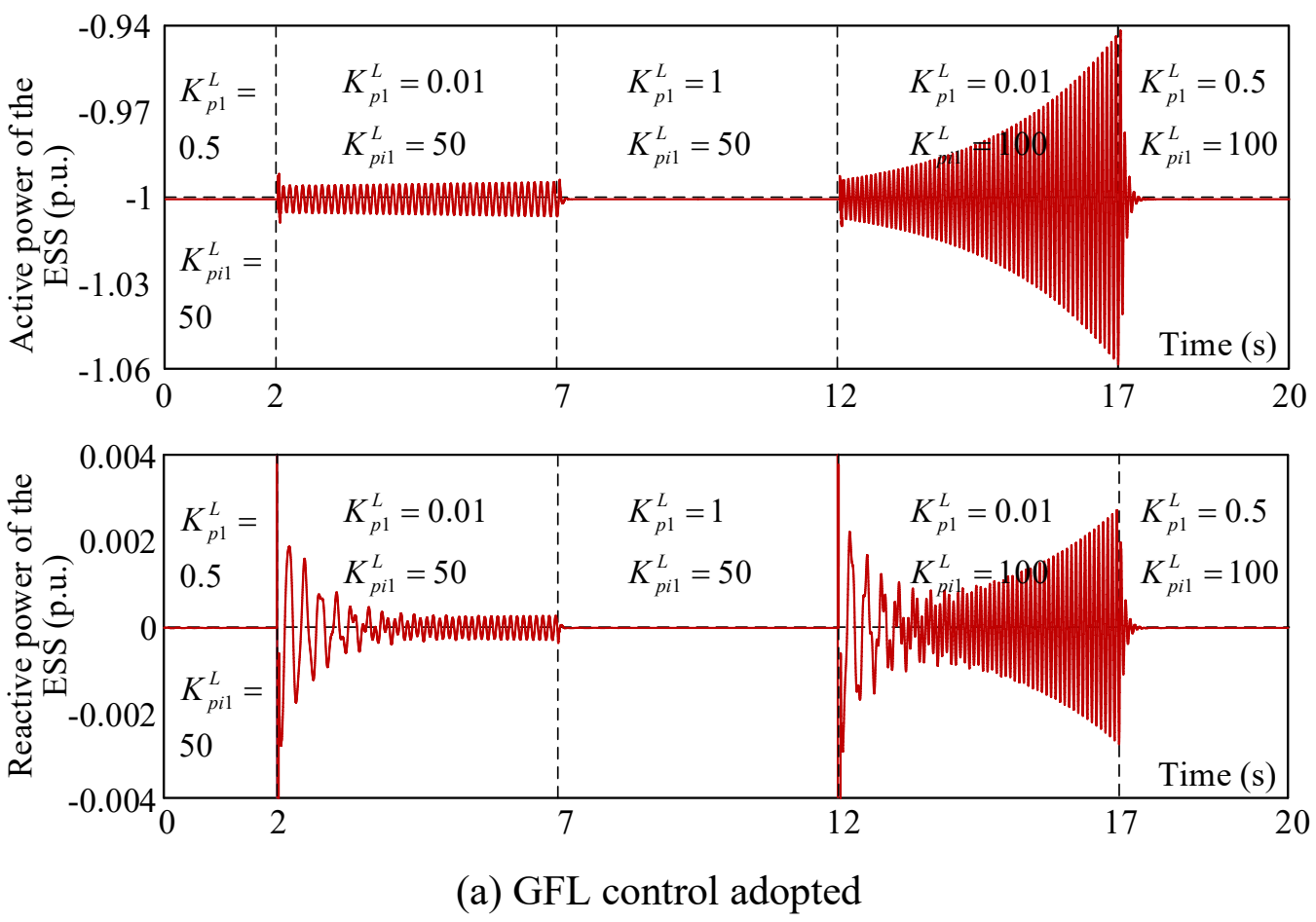

(a) GFL control adopted

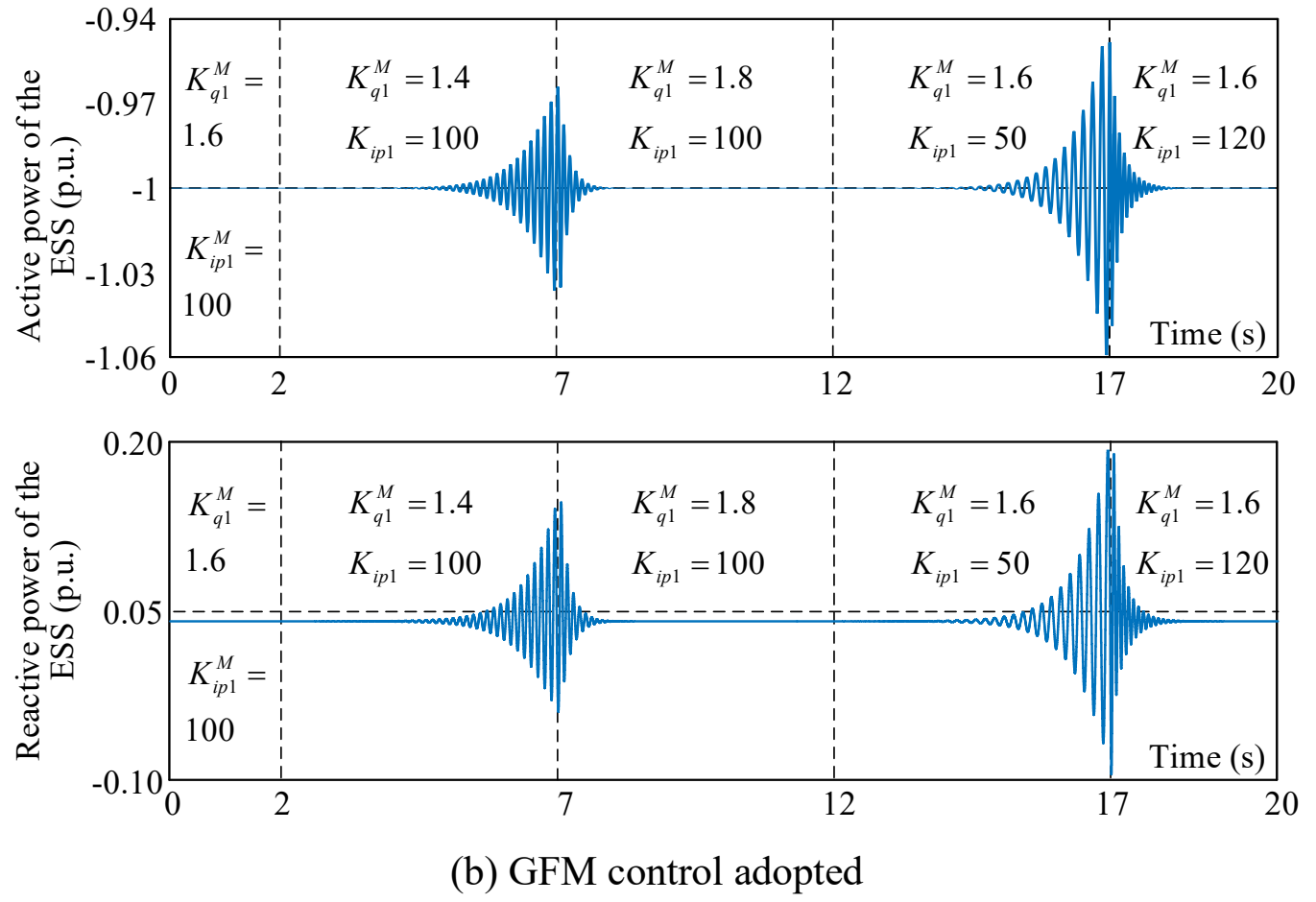

(b) GFM control adopted

Fig. 11. Time-domain simulation results when control parameters change.

The time-domain simulation results of Simulink are shown in Fig. 11, where control parameters change in a sequence. For GFL-CESSs (as shown in Fig. 11 (a)), the DVC loop parameters make a dominant imapct on stability. Increasing $K_{p1}^{L}$ enhances stability, while reducing $K_{p1}^{L}$ degrades damping performance. The integral gain $K_{pi1}^{L}$ predominantly oscillation frequency, with frequency rising proportionally to $K_{pi1}^{L}$. For GFM-CESSs (as shown in Fig. 11 (b)), stability is jointly governed by the AVC and PSC loops. Reducing $K_{q1}^{M}$ (AVC proportional gain) or $K_{ip1}$ (PSC integral gain) destabilizes the system, and instability occurs when $K_{q1}^{M}$ decreases to 1.4 or $K_{p1}^{L}$ decreases to 50 rad/s.

*D. Comparative Analysis of the Impact of d-q Control Loop Parameters on Oscillation Damping*

For GFL-CESSs, the dominant oscillation mode is $\lambda_s^L$ = -17.70 ± *j*56.86 rad/s, represented by the black circle in Fig. 12 (a). When the value of $K_{p1}^{L}$ changes from 0.5 to 0.1, $\lambda_s^L$ moves to the right side, so that the dynamic characteristics worsen and the damping reduces. Conversely, if $K_{p1}^{L}$ increases to 1, the damping improves. The trajectory of $\lambda_s^L$ variation is depicted by a blue line, and the terminal of oscillation mode is indicated by blue circles. Similarly, by changing the value of $K_{p1}^{L}$ in the range between 10 rad/s and 100 rad/s, the trajectory of the oscillation modes is illustrated by the red line. It can be observed that changing parameters of the *d*-axis loop has a significant effect on dynamic characteristics of the GFL-CESSs, where the damping is mainly affected by $K_{p1}^{L}$, and the frequency is mainly affected by $K_{pi1}^{L}$.

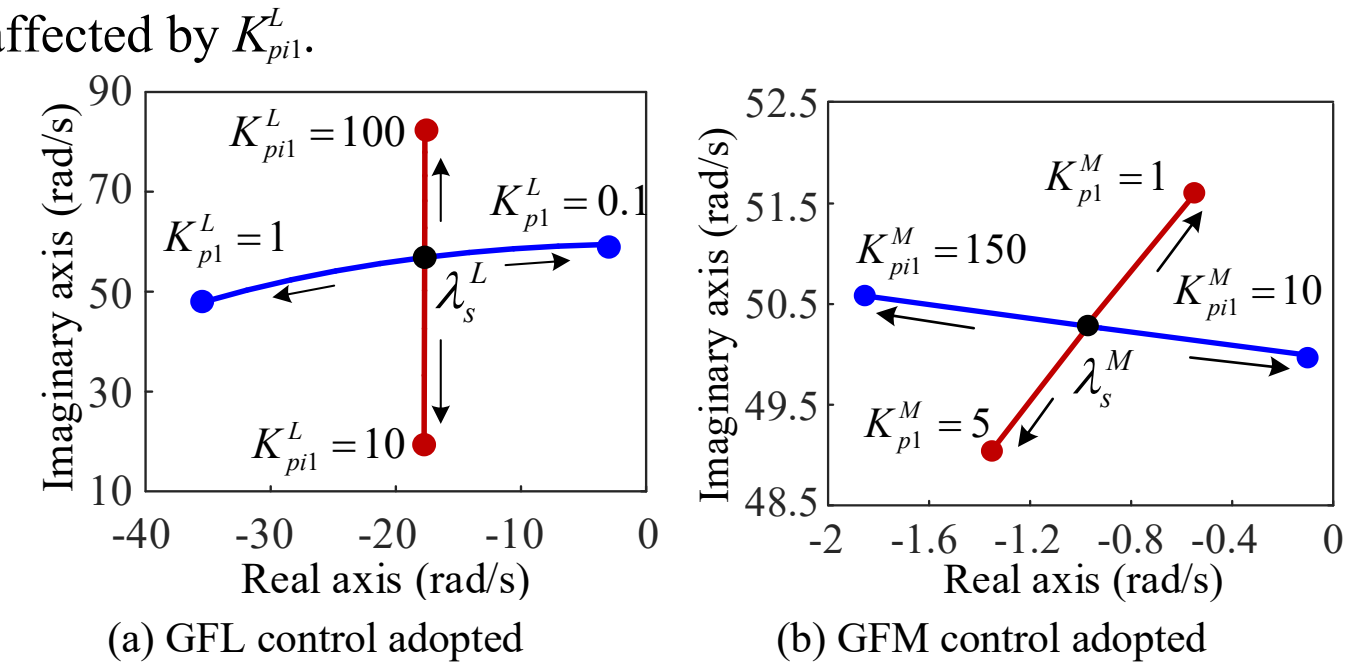

(a) GFL control adopted (b) GFM control adopted

Fig. 12. Trajectories of oscillation modes when *d*-axis control parameters vary.

For GFM-CESSs, the dominant oscillation mode is $\lambda_s^M$ = -0.97 ± *j*50.29 rad/s, represented by the black circle in Fig. 12 (b). When the value of $K_{p1}^{M}$ changes from 1 to 5, the oscillation damping varies from -0.55 rad/s to -1.35 rad/s, as depicted by a red line. When the value of $K_{pi1}^{M}$ changes from 10 rad/s to 150 rad/s, the damping varies from -0.1 rad/s to -1.85 rad/s depicted by a blue line. It can be seen that both parameters of the *d*-axis voltage control loop have a noticeable effect on the oscillation damping, which is different from that of GFL-CESSs.

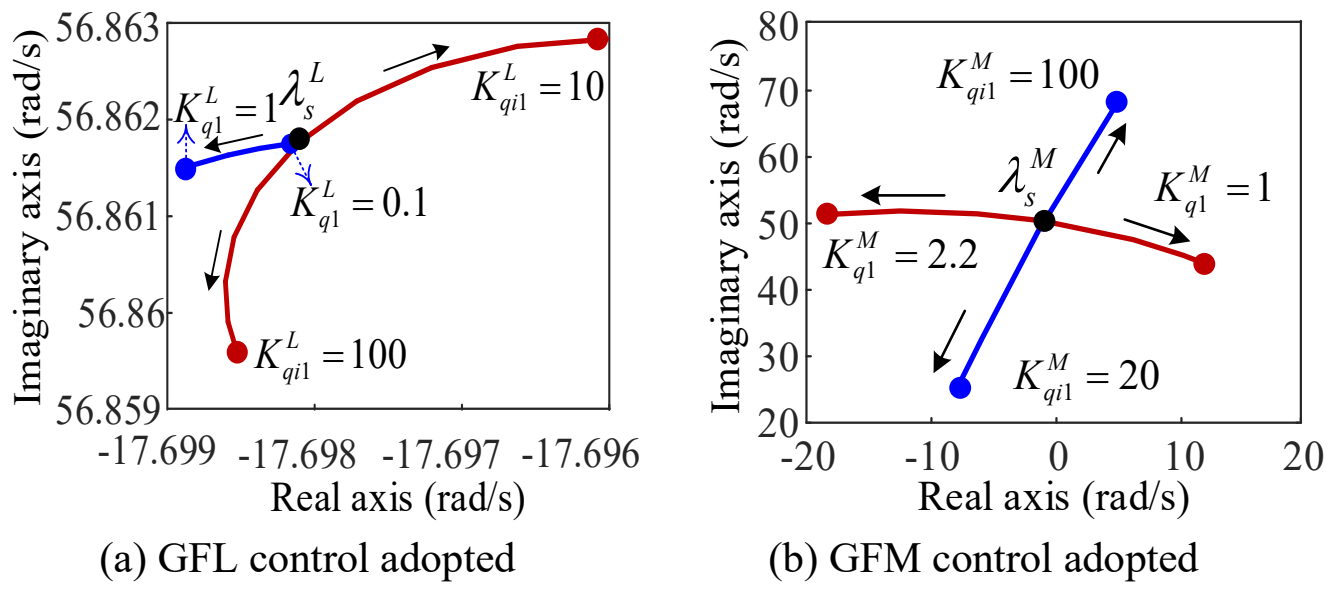

(a) GFL control adopted (b) GFM control adopted

Fig. 13. Trajectories of oscillation modes when *q*-axis control parameters vary.

Similarly, the same examinations are conducted for the *q*-axis outer control loops of GFL-CESSs and GFM-CESSs, with results illustrated in Fig. 13. From Fig. 13 (a), changes in *q*-axis control parameters of GFL-CESSs have negligible effects on oscillation mode locations, as the mode movement is minor. In contrast, Fig. 13(b) shows that parameter variations in the *q*-axis control loop of GFM-CESSs significantly change the oscillation mode, including both real and imaginary parts. Compared to GFL-CESSs, the *q*-axis voltage control loop of GFM-CESSs exhibits greater influence on the oscillation mode.

*E. Strong Dynamic Interaction between GFL-CESSs and GFM-CESSs under Hybrid Connections*

For hybrid connections of GFL-CESSs and GFM-CESSs, a total of five ESSs is considered. The initial control parameter values are as follows: GFL-CESSs: $K_{pl}^{L}$ = 0.2, $K_{ql}^{L}$ = 0.1, $K_{pil}^{L}$ = 80 rad/s, $K_{qil}^{L}$ = 10 rad/s. GFM-CESSs: $K_{p1}^{M}$ = 3.0, $K_{q1}^{M}$ = 1.0, $K_{pil}^{M}$ = 70 rad/s, $K_{qi1}^{M}$ = 50 rad/s, $K_{ip1}^{M}$ = 120 rad/s. The line reactance is given by $L_L$ = 0.1 p.u.

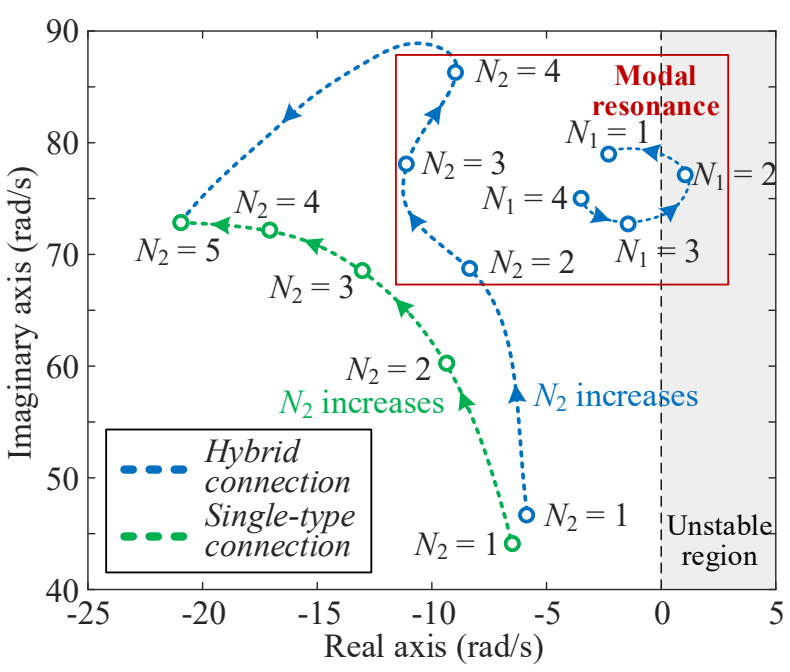

Fig. 14. Trajectories of oscillation modes when ESS number of GFM-CESSs increases.

Fig. 14 illustrates the trajectory of the dominant oscillation mode when the output power from each ESS is -1p.u. The ESS number of GFM-CESSs is varied from $N_2$ = 1 to $N_2$ = 5 (with the ESS number of GFL-CESSs decreasing from $N_1$ = 4 to $N_1$ = 0 accordingly). In Fig. 14, the oscillation mode for the case that only GFM-CESSs are connected (CESSs with single-type control) is indicated by green circles and lines as a comparison, while the case that both GFM-CESSs and GFL-CESSs are considered (CESSs with hybrid connection) is shown with blue circles and lines. It can be observed that when only GFM-CESSs are connected, the oscillation damping increases such that the system stability improves as $N_2$ increases.

However, for CESSs with hybrid connection where GFL-CESSs are also considered, the trajectory of dominant oscillation mode changes significantly compared to the single-type control case. When $N_2$ = 3, the dynamic interaction between GFL-CESSs and GFM-CESSs reaches its maximum, leading to instability in the power system. In this case, the oscillation modes are -11.13 + $j$78.1 rad/s and 0.99 + $j$77.34 rad/s. For comparison, when single-type control is adopted, the oscillation modes are -13.04 + $j$68.55 rad/s for GFM-CESSs and -3.736 + $j$77.22 rad/s for GFL-CESSs. Thus, when hybrid connection of GFL-CESSs and GFM-CESSs is considered, the wider shift in the oscillation mode of GFM-CESSs can lead to modal resonance with GFL-CESSs, which must be avoided in hybrid connections.

To resolve this issue, the ESS number of GFM-CESSs should be limited to narrow the region of their oscillation modes and avoid potential modal resonance risks. For example, when $N_2$ = 1, the dynamic interaction between GFM-CESSs and GFL-CESSs is weak. The oscillation modes are -5.93 + $j$46.68 rad/s and -3.513 + $j$75.0 rad/s for CESSs with hybrid connection, whereas the oscillation mode of GFM-CESSs is -6.49 + $j$44.12 rad/s for CESSs with single-type connection. In this case, the deviation in the oscillation mode as caused by the dynamic interaction is slight.

If the region of dominant oscillation mode of GFL-CESSs is identified, the control parameters of GFM-CESSs can be tuned so that their oscillation mode trajectory avoids the oscillation mode region of GFL-CESSs. In this study, the control parameters for GFM-CESSs are adjusted to $K_{q1}^{M}$ = 1.6 and $K_{ip1}^{M}$ = 50 rad/s, while keeping the remaining parameters unchanged. The oscillation mode trajectories are illustrated in Fig. 15. It is evident that the interaction between GFL-CESSs and GFM-CESSs is weakened compared to the original case illustrated in Fig. 14. The worst damping occurs when $N_2$ = 3, with a damping value of 2.91 rad/s, which ensures system stability.

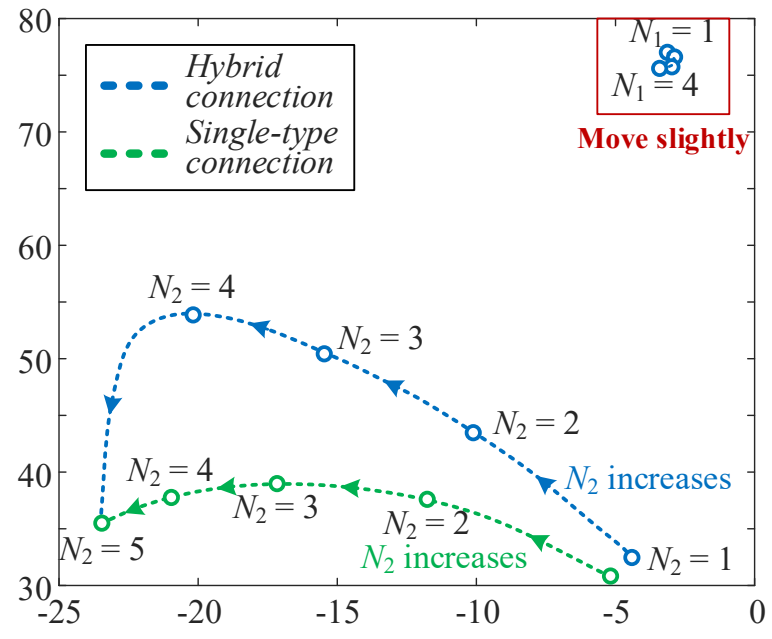

Fig. 15. Trajectories of oscillation modes when control parameters of GFM-CESSs change to $K_{q1}^{M}$ = 1.6 and $K_{ip1}^{M}$ = 50 rad/s.

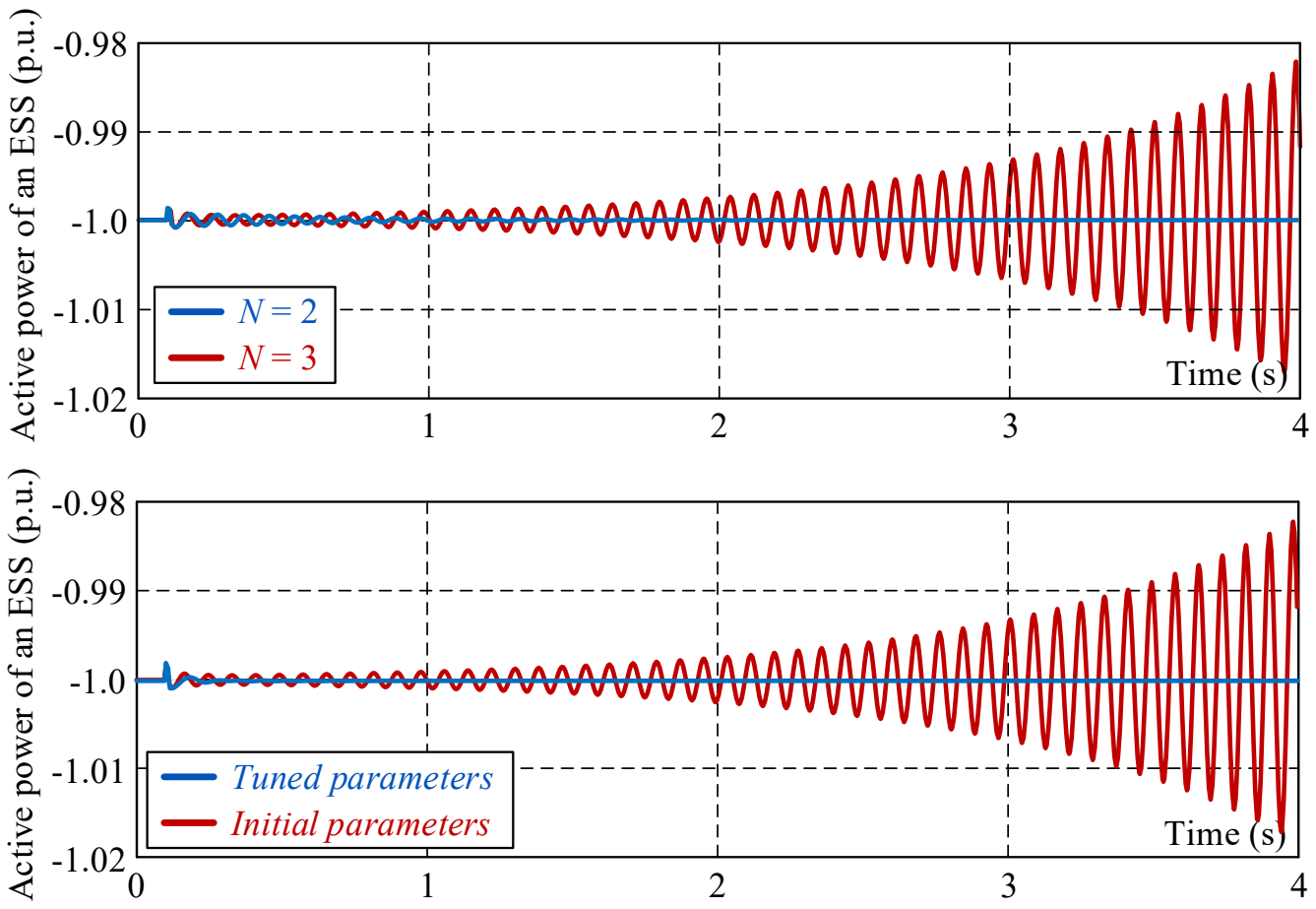

Fig. 16. Time-domain simulation results for ESSs with hybrid control.

For further confirmation, Fig. 16 presents the time-domain simulation results of the active power of the $1^{st}$ ESS. At 0.1 s, a disturbance of about 0.002 p.u. occurs in the active power and disappears after 0.01 s. As shown in Fig. 16 (a), the power system loses stability when $N_2$ = 3 but remains stable when $N_2$ = 1 and 2 when initial parameters are used. This confirms the instability risk for hybrid connections of GFL-CESSs and GFM-CESSs. Moreover, as demonstrated in Fig. 16 (b), proper tuning of control parameters can eliminate the instability and restore stable operation of the power system.

## VII. Conclusions

With the rapid advancement of energy storage technologies, the use of GFL or GFM control has sparked significant debate. This study evaluates the dynamic characteristics of CESSs under single-type and hybrid controls, providing actionable guidelines for control strategy selection for diverse operational scenarios.

For CESSs using single-type control, neither GFL-CESSs nor GFM-CESSs universally outperforms the other. The recommendation between GFL-CESSs and GFM-CESSs is scenario-dependent: GFL-CESSs are preferred under strong grid conditions, bidirectional power flow, and localized small-scale deployments. GFM-CESSs excel in weak grid connections and large-scale clustered applications.

For CESSs using hybrid control, increasing the ESS number of GFM-CESSs is not necessarily beneficial for damping improvement. The finding contrasts with traditional conclusions that encourage increasing the scale of GFM-CESS. A high ratio of GFM-CESSs can cause significant migration of the dominant oscillation mode, potentially leading to modal resonance with GFL-CESSs and decreasing stability. Therefore, the ratio of GFM-CESSs should be carefully determined and depends on the dynamics of GFL-CESSs.

If unstable oscillation occurs, parameter tuning is generally effective for mitigation. However, damping improvement is sensitive to different control loops for GFL-CESSs and GFM-CESSs. The dynamics of different control loops of GFM-CESSs are coupled. Therefore, parameter tuning should consider their internal coupling. In contrast, the control loops of GFL-CESSs are approximately decoupled, such that the dominant control loop can be identified and tuned for damping improvement.

Furthermore, the contributions of this study are applicable to centralized EVs, and those unrelated to bidirectional power flow are also applicable to centralized renewable energy clusters if the same scenario occurs.

## APPENDIX A

The dynamic equations of the filter are given by

$$\begin{aligned} sL_{fk}I_{dk} &= \omega_0 L_{fk} I_{qk} + V_{vdk} - V_{dk} \\ sL_{fk}I_{qk} &= -\omega_0 L_{fk} I_{dk} + V_{vqk} - V_{qk} \end{aligned}. \tag{A1}$$

The dynamic equations of the ICC loops are given by

$$\begin{aligned} \left(I_{dk}^{ref} - I_{dk}\right)\left(K_{cpk}^{L} + \frac{K_{cpik}^{L}}{s}\right) &= \omega_0 L_{fk} I_{qk} + V_{vdk} - V_{dk} \\ \left(I_{qk}^{ref} - I_{qk}\right)\left(K_{cqk}^{L} + \frac{K_{cqik}^{L}}{s}\right) &= -\omega_0 L_{fk} I_{dk} + V_{vqk} - V_{qk} \end{aligned}. \tag{A2}$$

Substituting (A2) into (A1), yields

$$\begin{aligned} sL_{fk}I_{dk} &= \left(K_{cpk}^{L} + \frac{K_{cpik}^{L}}{s}\right)\left(I_{dk}^{ref} - I_{dk}\right) \\ sL_{fk}I_{qk} &= \left(K_{cqk}^{L} + \frac{K_{cqik}^{L}}{s}\right)\left(I_{qk}^{ref} - I_{qk}\right) \end{aligned}. \tag{A3}$$

From the first equation of (A3), we obtain

$$I_{dk} = \frac{K_{cpk}^{L}s + K_{cpik}^{L}}{L_{fk}s^2 + K_{cpk}^{L}s + K_{cpik}^{L}} I_{dk}^{ref}. \tag{A4}$$

Similarly, from the second equation of (A3), we obtain

$$I_{qk} = \frac{K_{cqk}^{L}s + K_{cqik}^{L}}{L_{fk}s^2 + K_{cqk}^{L}s + K_{cqik}^{L}} I_{qk}^{ref}. \tag{A5}$$

## APPENDIX B

Considering the admittance matrix of AC network is in *x*-*y* coordinates. To interconnect the transfer functions in (5) and the admittance matrix, (5) needs to be transferred from *d*-*q* to *x*-*y* coordinates with

$$\begin{aligned} \begin{bmatrix} I_{dk} \\ I_{qk} \end{bmatrix} &= \begin{bmatrix} \cos\theta_{pk} & \sin\theta_{pk} \\ -\sin\theta_{pk} & \cos\theta_{pk} \end{bmatrix} \begin{bmatrix} I_{xk} \\ I_{yk} \end{bmatrix} \\ \begin{bmatrix} V_{dk} \\ V_{qk} \end{bmatrix} &= \begin{bmatrix} \cos\theta_{pk} & \sin\theta_{pk} \\ -\sin\theta_{pk} & \cos\theta_{pk} \end{bmatrix} \begin{bmatrix} V_{xk} \\ V_{yk} \end{bmatrix} \end{aligned}. \tag{B1}$$

By linearizing (B1), it yields

$$\begin{aligned} \begin{bmatrix} \Delta I_{xk} \\ \Delta I_{yk} \end{bmatrix} &= \mathbf{T}\begin{bmatrix} \Delta I_{dk} \\ \Delta I_{qk} \end{bmatrix} + \dot{\mathbf{T}}\begin{bmatrix} I_{dk,0} & 0 \\ 0 & I_{qk,0} \end{bmatrix}\begin{bmatrix} \Delta\theta_{pk} \\ \Delta\theta_{pk} \end{bmatrix} \\ \begin{bmatrix} \Delta V_{dk} \\ \Delta V_{qk} \end{bmatrix} &= \mathbf{T}_{xy2dq}\begin{bmatrix} \Delta V_{xk} \\ \Delta V_{yk} \end{bmatrix} + \dot{\mathbf{T}}_{xy2dq}\begin{bmatrix} V_{xk,0} & 0 \\ 0 & V_{yk,0} \end{bmatrix}\begin{bmatrix} \Delta\theta_{pk} \\ \Delta\theta_{pk} \end{bmatrix}, \\ \begin{bmatrix} \Delta\theta_{pk} \\ \Delta\theta_{pk} \end{bmatrix} &= p_{\theta k}(s)\mathbf{T}_{th}\begin{bmatrix} \Delta V_{xk} \\ \Delta V_{yk} \end{bmatrix} \end{aligned} \tag{B2}$$

where

$$\mathbf{T} = \mathbf{T}_{xy2dq}^{-1} = \begin{bmatrix} \cos\theta_{pk,0} & -\sin\theta_{pk,0} \\ \sin\theta_{pk,0} & \cos\theta_{pk,0} \end{bmatrix},$$

$$\dot{\mathbf{T}} = \begin{bmatrix} -\sin\theta_{pk,0} & -\cos\theta_{pk,0} \\ \cos\theta_{pk,0} & -\sin\theta_{pk,0} \end{bmatrix} \quad \dot{\mathbf{T}}_{xy2dq} = \begin{bmatrix} -\sin\theta_{pk,0} & \cos\theta_{pk,0} \\ -\cos\theta_{pk,0} & -\sin\theta_{pk,0} \end{bmatrix},$$

$$\mathbf{T}_{th} = \frac{1}{V_{k,0}^2}\begin{bmatrix} -V_{yk,0} & V_{xk,0} \\ -V_{yk,0} & V_{xk,0} \end{bmatrix}, \quad \mathbf{T}_{xy2dq} = \begin{bmatrix} \cos\theta_{pk,0} & \sin\theta_{pk,0} \\ -\sin\theta_{pk,0} & \cos\theta_{pk,0} \end{bmatrix},$$

$$p_{\theta k}(s) = \frac{\Delta\theta_{pk}}{\Delta\theta_k} = -\frac{V_{k,0}\left(K_{\theta pk}^{L}s + K_{\theta ik}^{L}\right)}{s^2 - V_{k,0}\left(K_{\theta pk}^{L}s + K_{\theta ik}^{L}\right)}.$$

Substituting (B2) into (4), and eliminating $\Delta I_{dk}$, $\Delta I_{qk}$, $\Delta V_{dk}$, and $\Delta V_{qk}$, the following transfer function can be derived:

$$\begin{aligned} \mathbf{Y}_{xyk}^{L}(s) = \mathbf{T}\mathbf{Y}_{dqk}^{L}(s)&\left(\mathbf{T}_{xy2dq} + \dot{\mathbf{T}}_{xy2dq}\begin{bmatrix} V_{xk,0} & 0 \\ 0 & V_{yk,0} \end{bmatrix} p_{\theta k}(s)\mathbf{T}_{th}\right) \\ &+ \dot{\mathbf{T}}\begin{bmatrix} I_{dk,0} & 0 \\ 0 & I_{qk,0} \end{bmatrix} p_{\theta k}(s)\mathbf{T}_{th} \end{aligned}. \tag{B3}$$

## APPENDIX C

Equation (A2) can be rewritten as:

$$\begin{aligned} \Delta\mathbf{I}_{dq} &= \mathbf{T}_{xy2dq}\Delta\mathbf{I}_{xy} + \dot{\mathbf{T}}_{xy2dq}\mathbf{I}_{xy0}{}^{T}\Delta\theta_k \\ \Delta\mathbf{V}_{dq} &= \mathbf{T}_{xy2dq}\Delta\mathbf{V}_{xy} + \dot{\mathbf{T}}_{xy2dq}\mathbf{V}_{xy0}{}^{T}\Delta\theta_k \end{aligned}, \tag{C1}$$

where $\mathbf{I}_{xy0} = Diag(I_{xk,0}, I_{yk,0})$, $\mathbf{V}_{xy0} = Diag(V_{xk,0}, V_{yk,0})$.

The variation in $\Delta\theta_k$ is composed of two parts as below:

$$\Delta\theta_k = \Delta\theta_{k1} + \Delta\theta_{k2}. \tag{C2}$$

In (C2), $\Delta\theta_{k1}$ is caused by the dynamic progress of angle control loop, which can be written as

$$\Delta\theta_{k1} = K_{ipk}(s)\Delta P_{tk}, \tag{C3}$$

where $K_{ipk}(s) = -K_{ipk}s^{-1}$, $\Delta P_{tk} = \mathbf{V}_{xy0}\Delta\mathbf{I}_{xy} + \mathbf{I}_{xy0}\Delta\mathbf{V}_{xy}$.

In (C2), $\Delta\theta_{k2}$ is caused by the variation of angle of $V_k$, which can be written as

$$\Delta\theta_{k2} = \mathbf{T}_{th}^{M}\Delta\mathbf{V}_{xy}, \tag{C4}$$

where $\mathbf{T}_{th}^{M} = -\begin{bmatrix} -\frac{V_{yk,0}}{\left|V_{k,0}\right|^2} & \frac{V_{xk,0}}{\left|V_{k,0}\right|^2} \end{bmatrix}$.

The linearized equations of (6) and (7) are as follows:

$$\Delta \mathbf{I}_{dq} = \mathbf{G}_{idqk}^{M}(s)\mathbf{K}_{vdqk}(s)\Delta \mathbf{V}_{dq}\,, \tag{C5}$$

where $\mathbf{G}_{idqk}^{M}(s) = Diag(G_{idk}^{M}(s), G_{iqk}^{M}(s))$,

$\mathbf{K}_{vdqk}(s) = Diag(-(K_{pk}^{M} + K_{pik}^{M}s^{-1}), -(K_{qk}^{M} + K_{qik}^{M}s^{-1}))$.

Substituting (C2) to (C4) into (C1), and eliminating $\Delta\theta_k$ yields

$$\begin{aligned}
&\Delta \mathbf{V}_{dq} = \dot{\mathbf{T}}_{xy2dq}\mathbf{V}_{xy0}{}^{T} K_{ipk}(s)\mathbf{V}_{xy0}\Delta \mathbf{I}_{xy} + \\
&\mathbf{T}_{xy2dq}\Delta \mathbf{V}_{xy} + \dot{\mathbf{T}}_{xy2dq}\mathbf{V}_{xy0}{}^{T}\left(K_{ipk}(s)\mathbf{I}_{xy0} + \mathbf{T}_{th}^{M}\right)\Delta \mathbf{V}_{xy} \\
&\Delta \mathbf{I}_{dq} = \mathbf{T}_{xy2dq}\Delta \mathbf{I}_{xy} + \dot{\mathbf{T}}_{xy2dq}\mathbf{I}_{xy0}{}^{T} K_{ipk}(s)\mathbf{V}_{xy0}\Delta \mathbf{I}_{xy} \\
&+\dot{\mathbf{T}}_{xy2dq}\mathbf{I}_{xy0}{}^{T}\left(K_{ipk}(s)\mathbf{I}_{xy0} + \mathbf{T}_{th}^{M}\right)\Delta \mathbf{V}_{xy}
\end{aligned}. \tag{C6}$$

Substituting (C5) into (C6), and eliminating $\Delta\mathbf{V}_{dq}$ and $\Delta\mathbf{I}_{dq}$, yields

$$\begin{aligned}
&\mathbf{G}_{idqk}^{M}(s)\mathbf{K}_{vdqk}(s)\dot{\mathbf{T}}_{xy2dq}\mathbf{V}_{xy0}{}^{T} K_{ipk}(s)\mathbf{V}_{xy0}\Delta \mathbf{I}_{xy} + \\
&\mathbf{G}_{idqk}^{M}(s)\mathbf{K}_{vdqk}(s)\mathbf{T}_{xy2dq}\Delta \mathbf{V}_{xy} + \\
&\mathbf{G}_{idqk}^{M}(s)\mathbf{K}_{vdqk}(s)\dot{\mathbf{T}}_{xy2dq}\mathbf{V}_{xy0}{}^{T}\left(K_{ipk}(s)\mathbf{I}_{xy0} + \mathbf{T}_{xy}\right)\Delta \mathbf{V}_{xy} \\
&= \mathbf{T}_{xy2dq}\Delta \mathbf{I}_{xy} + \dot{\mathbf{T}}_{xy2dq}\mathbf{I}_{xy0}{}^{T} K_{ipk}(s)\mathbf{V}_{xy0}\Delta \mathbf{I}_{xy} \\
&+\dot{\mathbf{T}}_{xy2dq}\mathbf{I}_{xy0}{}^{T}\left(K_{ipk}(s)\mathbf{I}_{xy0} + \mathbf{T}_{xy}\right)\Delta \mathbf{V}_{xy}
\end{aligned}. \tag{C7}$$

According to (C7), the transfer function of the VSC with GFM control can be obtained as

$$\mathbf{Y}_{xyk}^{M}(s) = \frac{\mathbf{G}_{idqk}^{M}(s)\mathbf{K}_{vdqk}(s)\mathbf{T}_{xy2dq} - \mathbf{T}_{s}\left(K_{ipk}(s)\mathbf{I}_{xy0} + \mathbf{T}_{th}^{M}\right)}{\mathbf{T}_{xy2dq} + \mathbf{T}_{s}K_{ipk}(s)\mathbf{V}_{xy0}}, \tag{C8}$$

where $\mathbf{T}_{s} = \dot{\mathbf{T}}_{xy2dq}\mathbf{I}_{xy0}{}^{T} - \mathbf{G}_{idqk}^{M}(s)\mathbf{K}_{vdqk}(s)\dot{\mathbf{T}}_{xy2dq}\mathbf{V}_{xy0}{}^{T}$.

## REFERENCES

[1] L. Xu, X. Luo, Y. Wen et al., "Energy management of hybrid power ship system using adaptive moth flame optimization based on multi-populations," IEEE Trans. Power Syst., vol. 39, no. 1, pp. 1711–1727, Jan. 2024.

[2] M. Rouholamini, C. Wang, H. Nehrir et al., "A review of modeling, management, and applications of grid-connected Li-ion battery storage systems," IEEE Trans. Smart Grid, vol. 13, no. 6, pp. 4505–4524, Nov. 2022.

[3] F. Salehi, A. Golshani, I. B. M. Matsuo, P. Dehghanian, M. Aghazadeh Tabrizi, W.-J. Lee, "On mitigation of sub-synchronous control interactions in hybrid generation resources," IEEE Trans. Ind. Informat., vol. 18, no. 7, pp. 4372–4382, Jul. 2022.

[4] DOE Global Energy Storage Database. [Online]. Available: https://sandia.gov/ess-ssl/gesdb/public/statistics.html [Accessed: Dec. 2024].

[5] M. Ghazavi Dozein, O. Gomis-Bellmunt, P. Mancarella, "Simultaneous provision of dynamic active and reactive power response from utility-scale battery energy storage systems in weak grids," IEEE Trans. Power Syst., vol. 36, no. 6, pp. 5548–5557, Nov. 2021.

[6] K. Saleem, Z. Ali, K. Mehran, "A single-phase synchronization technique for grid-connected energy storage system under faulty grid conditions," IEEE Trans. Power Electron., vol. 36, no. 10, pp. 12019–12032, Oct. 2021.

[7] M. A. K. Magableh, A. A. A. Radwan, Y. A.-R. I. Mohamed, "Dynamic analysis and stability enhancement of a weak grid-tied hybrid system integrating PV, full-scale wind turbine, and battery storage," IEEE Open J. Ind. Appl., vol. 6, pp. 325–349, 2025.

[8] R. Hemmati, A. Neda, "Optimal control strategy on battery storage systems for decoupled active-reactive power control and damping oscillations," J. Energy Storage, vol. 13, pp. 24–34, Oct. 2017.

[9] J. Chen, J. Chen, "Stability analysis and parameters optimization of islanded microgrid with both ideal and dynamic constant power loads," IEEE Trans. Ind. Electron., vol. 65, no. 4, pp. 3263–3274, Apr. 2018.

[10] W. Du, W. Dong, Y. Wang, H. Wang, "Small-disturbance stability of a wind farm with virtual synchronous generators under the condition of weak grid connection," IEEE Trans. Power Syst., vol. 36, no. 6, pp. 5500–5511, Nov. 2021.

[11] Y. Hirase, K. Abe, K. Sugimoto et al., "A novel control approach for virtual synchronous generators to suppress frequency and voltage fluctuations in microgrids," Appl. Energy, vol. 210, pp. 699–710, Jan. 2018.

[12] Y. Liu, Y. Chen, H. Xin et al., "System strength constrained grid-forming energy storage planning in renewable power systems," IEEE Trans. Sustain. Energy, vol. 16, no. 2, pp. 981–994, Apr. 2025.

[13] F. Zhao, X. Wang, Z. Zhou, Ł. Kocewiak, J. R. Svensson, "Comparative study of battery-based STATCOM in grid-following and grid-forming modes for stabilization of offshore wind power plant," Electr. Power Syst. Res., vol. 212, p. 108449, 2022.

[14] S. Jiang, G. Konstantinou, "Generalized impedance model and interaction analysis for multiple grid-forming and grid-following converters," Electr. Power Syst. Res., vol. 214, pt. B, p. 108912, 2023.

[15] J. Alipoor, Y. Miura, T. Ise, "Power system stabilization using virtual synchronous generator with alternating moment of inertia," IEEE J. Emerg. Sel. Topics Power Electron., vol. 3, no. 2, pp. 451–458, Jun. 2015.

[16] X. Gao, D. Zhou, A. Anvari-Moghaddam, F. Blaabjerg, "Stability analysis of grid-following and grid-forming converters based on state-space model," in Proc. IPEC-Himeji 2022 ECCE Asia, Himeji, Japan, 2022, pp. 422–428.

[17] F. Zhao, X. Wang, T. Zhu, "Power dynamic decoupling control of grid-forming converter in stiff grid," IEEE Trans. Power Electron., vol. 37, no. 8, pp. 9073–9088, Aug. 2022.

[18] W. Du, Q. Fu, H. F. Wang, "Power system small-signal angular stability affected by virtual synchronous generators," IEEE Trans. Power Syst., vol. 34, no. 4, pp. 3209–3219, Jul. 2019.

[19] A. A. Radwan, Y. A.-R. I. Mohamed, "Analysis and active-impedance-based stabilization of voltage-source-rectifier loads in grid-connected and isolated microgrid applications," IEEE Trans. Sustain. Energy, vol. 4, no. 3, pp. 563–576, Jul. 2013.

[20] M. Cespedes, L. Xing, J. Sun, "Constant-power load system stabilization by passive damping," IEEE Trans. Power Electron., vol. 26, no. 7, pp. 1832–1836, Jul. 2011.

[21] Q. Fu, W. Du, X. Chen, H. F. Wang, X. Xiao, "Dynamic analysis of energy storage integrated systems considering bidirectional power flow and different control loops of energy storages," J. Energy Storage, vol. 86, p. 111171, 2024.

[22] X. Wang, M. G. Taul, H. Wu, Y. Liao, F. Blaabjerg, L. Harnefors, "Grid-synchronization stability of converter-based resources—An overview," IEEE Open J. Ind. Appl., vol. 1, pp. 115–134, 2020.

[23] M. Li, Y. Wang, W. Hu et al., "Unified modeling and analysis of dynamic power coupling for grid-forming converters," IEEE Trans. Power Electron., vol. 37, no. 2, pp. 2321–2337, Feb. 2022.

[24] F. Zhao, X. Wang, Z. Zhou et al., "Control interaction modeling and analysis of grid-forming battery energy storage system for offshore wind power plant," IEEE Trans. Power Syst., vol. 37, no. 1, pp. 497–507, Jan. 2022.

[25] C. Collados-Rodriguez, M. Cheah-Mane, E. Prieto-Araujo, O. Gomis-Bellmunt, "Stability analysis of systems with high VSC penetration: Where is the limit?," IEEE Trans. Power Del., vol. 35, no. 4, pp. 2021–2031, Aug. 2020.

[26] V. Fernão Pires, E. Romero-Cadaval, D. Vinnikov, I. Roasto, J. F. Martins, "Power converter interfaces for electrochemical energy storage systems – A review, " Energy Convers. Manage., vol. 86, pp. 453–475, 2014.

[27] X. Lin, R. Zamora, "Controls of hybrid energy storage systems in microgrids: Critical review, case study and future trends," J. Energy Storage, vol. 47, p. 103884, 2022.

[28] J. Shair, H. Z. Li, J. B. Hu et al., "Stability improvement of microgrids in the presence of constant power loads," Int. J. Electr. Power Energy Syst., vol. 36, pp. 442–456, Mar. 2018.

[29] Q. Fu, W. Du, H. F. Wang, X. Xiao, "Analysis of subsynchronous oscillation caused by multiple VSCs with different dynamics under strong grid connections," IEEE Trans. Sustain. Energy, vol. 14, no. 4, pp. 2364–2375, Oct. 2023.

[30] Q. Fu, C. Dai, S. Bu, C. Y. Chung, "Integrating power electronics-based energy storages to power systems: A review on dynamic modeling, analysis, and future challenges," Renew. Sustain. Energy Rev., vol. 213, p. 115460, 2025.

[31] H. Yuan, X. Yuan, J. Hu, "Modeling of grid-connected VSCs for power system small-signal stability analysis in DC-link voltage control timescale," IEEE Trans. Power Syst., vol. 32, no. 5, pp. 3981–3991, Sep. 2017.

[32] W. Du, Q. Fu, H. Wang, "Damping torque analysis of DC voltage stability of an MTDC network for the wind power delivery," IEEE Trans. Power Del., vol. 35, no. 1, pp. 324–338, Feb. 2020.

[33] Q. Fu, W. Du, H. F. Wang, "Analysis of harmonic oscillations caused by grid-connected VSCs," IEEE Trans. Power Del., vol. 36, no. 2, pp. 1202–1210, Apr. 2021.

[34] W. Du, Q. Fu, H. F. Wang, "Open-loop modal coupling analysis for a multi-input multi-output interconnected MTDC/AC power system," IEEE Trans. Power Syst., vol. 34, no. 1, pp. 246–256, Jan. 2019.

[35] P. Giroux, G. Sybille, C. Osorio, S. Chandrachood, “Average model of a 100-kW grid-connected PV array,” MathWorks. [Online]. Available: https://ww2.mathworks.cn/help/sps/ug/average-model-of-a-100-kw-grid-connected-pv-array.html. [Accessed: Dec. 2024].